\documentclass[aps,twocolumn,pra,groupaddress,showpacs,preprintnumbers,amsmath,amssymb]{revtex4-1}
\newcommand{\Tr}{\operatorname{Tr}}
\newcommand{\Det}{\operatorname{Det}}
\newcommand{\G}{{\bf G}}
\newcommand{\M}{{\bf M}}

\newcommand{\K}{{\bf k}}
\newcommand{\Kp}{{\bf k'}}
\newcommand{\Q}{{\bf q}}
\newcommand{\sumk}{{\sum_{ik_n,{\bf k}}}}
\newcommand{\sumq}{{\sum_{iq_l,{\bf q}}}}

\newcommand{\mut}{{\tilde \mu}}

\ifx\pdfoutput\undefined
\usepackage{graphicx}
\usepackage[dvips]{color}
\usepackage{epsfig}
\else
\usepackage[pdftex]{graphicx}
\usepackage{epstopdf}
\fi
\usepackage{dcolumn}
\usepackage{bm}
\begin{document}
\title{BCS-BEC crossover in an optical lattice}
\author{Parag Ghosh}
\affiliation{Department of Physics, University of Illinois, Urbana-Champaign, Illinois 61801, USA}
\date{\today}
\begin{abstract}
We model fermions with an attractive interaction in an optical lattice
with a single-band Hubbard model away from half-filling with on-site
attraction $U$ and nearest neighbor hopping $t$. Our goal is to understand the crossover from BCS
superfluidity in the weak attraction limit
to the BEC of molecules in the strong attraction limit, with
particular emphasis on how this crossover in an optical lattice
differs from the much better studied continuum problem. We use a large-$N$
theory with Sp($2N$) symmetry to study the fluctuations beyond mean
field theory. At $T=0$, we calculate across the crossover various
observables, including chemical potential, gap, ground state energy,
speed of sound and compressibility. The superfluid density $n_s$ is
found to have non-trivial $U/t$ dependence in this lattice system. We
show that the transition temperature $T_c$ scales with the energy gap
in the weak coupling limit but crosses over to a $t^2/U$ scaling in
the BEC limit, where phase fluctuations controlled by $n_s$
determine $T_c$. We also find, quite contrary to our expectations, that in the strong coupling limit, the large-$N$ theory gives qualitatively wrong trends for compressibility. A comparison with a simple Hartree shifted BCS theory, which takes into account both pairing and Hartree
shifts, and correctly recovers the atomic limit and the right qualitative trend for compressibility, reveals that the large-$N$ theory on the lattice, although considers a larger number of diagrams, is in fact inferior to the simpler Hartree shifted BCS theory. The failure of the large-$N$ approach is explained by noting (i) the importance of Hartree shift in lattice problems, and (ii) inability of the large-$N$ approach to treat particle-particle and particle-hole channels at equal footing at the saddle point level.
\end{abstract}
\maketitle
\section{Introduction}
The problem of the BCS-BEC crossover of strongly interacting fermions has been well studied in the continuum both theoretically~\cite{leggett80, eagles, nsr, BECreview, sademelo, rsr, engelbrecht97, Haussman, vsr, Sachdev, Carlson, Pieri-Strinati} and experimentally~\cite{jochim, jin, ketterle}. The system smoothly interpolates between a BCS state of loosely bound Cooper pairs to a Bose-Einstein condensate of tightly bound diatomic molecules. The Leggett-BCS mean field theory gives qualitatively correct physics at $T=0$ across the crossover and methods like functional integral formalism~\cite{sademelo, rsr, engelbrecht97}, self-consistent approximations~\cite{Haussman}, large-$N$ expansions~\cite{vsr, Sachdev} and quantum Monte Carlo~\cite{Carlson, Pieri-Strinati} have been used to find quantitative corrections. Experiments have demonstrated the condensation of molecules in the BEC limit~\cite{jochim, jin} and the superfluidity of the system across resonance has been observed~\cite{ketterle}.

The inclusion of a lattice in the system leads to several qualitative differences with the continuum.
One of the key features distinguishing a lattice system from the continuum is the dependence on interaction strength and filling fraction of the superfluid stiffness of the gas even at $T=0$. This is in contrast to the continuum case, where the $T=0$ superfluid stiffness is fixed by the particle mass and density due to Galilean invariance. Consequently, when phase fluctuations play a dominant role in the loss of phase coherence and the superfluid stiffness sets the scale for transition temperature~\cite{emery-kivelson}, the above mentioned difference between the lattice and continuum becomes explicit. A second difference, which is not entirely unrelated to the previous point, that arises between the continuum and the lattice is regarding the effective mass of the bound pairs in the BEC limit. In the continuum, the mass of the bound pair in the BEC limit is simply twice the mass of the fermions and hence does not scale with the coupling strength. In contrast, the effective mass of the bosons on the lattice becomes increasing larger with the strength of the coupling. This is due to the fact that the bosons on the lattice can only move around by virtual ionization, and hence the corresponding hopping matrix element for the bosons, calculated within a simple perturbation theory, has an energy denominator equal to the coupling strength. Consequently, the boson mass which is inversely proportional to the hopping becomes larger with the strength of coupling. Thirdly, on a bipartite lattice there is a particle-hole (p-h) transformation that puts additional constraints on thermodynamics. Finally, there is an emergence on a lattice of a Charge Density Wave order at half-filling that competes with the superfluid (pairing) order. This new order arises because at half-filling the lattice Hamiltonian has a higher SU(2) symmetry in the spin space that is spontaneously broken. Our primary objective for studying attractive fermionic atoms on a lattice is therefore to understand how the broken translational invariance affects various physical quantities across the crossover. Moreover, there is a growing interest in performing experiments with ultra-cold fermionic clouds of both $^{40}$K~\cite{eth} and $^6$Li~\cite{mit} atoms in optical lattices, and although the entropy in the current experiments needs to be reduced by a factor of 3 or higher in order to reduce the temperature to below $T_c$ \cite{paiva}, we believe that in future these experiments should be able to test the findings of our current work.

The paper is organized as follows: In section II we introduce the Hamiltonian and discuss the p-h constraints on a lattice. In section III we discuss a Hartree shifted BCS (HBCS) theory that respects the lattice p-h constraints. In section IV we next outline a diagrammatic approach to include the effects of quantum fluctuations on top of the HBCS theory and discuss the problem in this approach. Next, in section V we develop a large-$N$ formalism that we use in the rest of the paper. We discuss the $T=0$ results for different properties of the system within the large-$N$ approach. We conclude this section with a comparison between the HBCS and large-$N$, and show that the former theory gives a more correct account of the chemical potential and the compressibility at $T=0$. In section V we calculate the zero temperature superfluid stiffness. In Section VI we outline the calculation and results for the critical temperature. We conclude in section VIII.

\section{Hamiltonian and particle-hole constraints}
In this section we first introduce the Hamiltonian that describes two kinds of fermions in a lattice and calculate the strength of interactions for which a two-particle bound state appears (unitarity condition). We next derive a set of p-h constraints imposed on the thermodynamics and develop a simple HBCS theory that respects these constraints.

\subsection{Hamiltonian}
The study of the BCS-BEC crossover in the absence of an optical lattice uses the divergence of the scattering length near a Feshbach resonance to tune the strength of the interactions between the fermions. Although this same technique has been applied to fermions in optical lattices~\cite{mit}, the Hamiltonian that describes this system near resonance is poorly understood. 
This is due to the inherent multi-band nature of the system when the (continuum) scattering length between the atoms diverges \citep{Diener-Ho-lattice}.
We do not have a separation of energy scales that would allow us to study an effective Hamiltonian in a single band.
However, as we will show below, the lattice strongly modifies the scattering properties of fermions restricted to the lowest band, to the point that it takes a finite amount of on-site interaction to form a (molecular) bound state. Thus, a Feshbach resonance is not needed to achieve a unitary gas in a lattice.

The Hamiltonian we will study is the single-band attractive Hubbard Hamiltonian:
\begin{equation}
H=-t\sum_{\langle i,j \rangle, \sigma} (c^{\dagger}_{i\sigma}c_{j\sigma}+c^{\dagger}_{j\sigma}c_{i\sigma}) - U \sum_i
n_{i\uparrow}n_{i\downarrow}-\mu\sum_{i, \sigma}n_{i\sigma}.
\label{ham}
\end{equation}
Here $c_{j\sigma}$ is the fermion annihilation operator at site $j$, the pseudo-spin index $\sigma=\uparrow, \downarrow$ represent the
two-hyperfine states, $t$ is the hopping matrix element between adjacent sites and the summation indices ${\langle
i,j \rangle}$ represent sums over nearest-neighbor sites. The on-site attractive coupling is given by $-U$ with
$U>0$, and it is assumed that both the hopping $t$ and $U$ are much smaller than the inter-band gap. Finally, $n_{i\sigma} = c^\dagger_{i\sigma} c_{i\sigma}$ is the number operator at site $i$ of fermions with spin $\sigma$, and $\mu$ is the chemical potential. 
For simplicity, we will study homogeneous systems; i.e. we neglect the effects of the (typically harmonic) external trapping potential, which can eventually be included using a local density approximation. Throughout the paper, we have set $\hbar=k_B=1$ and we shall use the convention that all 3-momenta sums are summed over the first Brillouin zone and then divided by the total number of lattice sites. 

The scattering amplitude between fermions in the lattice can be obtained by summing up all possible interaction events of fermions with the dispersion relation obtained from the kinetic energy in (\ref{ham}), $\epsilon_{\K}=-2t[\cos(k_x a)+\cos(k_y a)+\cos(k_z a)-3]$ (which we conventionally measure from the bottom of the band), where $a$ is the lattice constant. The scattering amplitude can be calculated as $f = (m/4\pi) \, \Gamma(0,0)$ where 
$\Gamma(\Q, \omega)=U/(1+U \Pi(\Q, \omega))$ is the four-point vertex function for a pair of fermions of mass $m$ with center of mass momentum $\Q$ and $\Pi(\Q, \omega)$ is the corresponding polarization, which in our case (and in the limit $T\rightarrow 0$) is of the form
\begin{equation}
\Pi(\Q, \omega)=\int_{{\textup BZ}} {d{\K}\over(2\pi)^3} \frac{1}{\omega+\epsilon_{\Q/2+\K}+\epsilon_{\Q/2-\K}}, 
\end{equation}
where the integration is over the Brillouin zone. We can now see that the condition for a diverging scattering amplitude (i.e. unitarity) in the lattice is~\cite{svistunov}:
\begin{equation}
\frac{1}{U^*}=-\Pi({\bf 0}, 0)=-\sum_{\K} \frac{1}{2\epsilon_{\K}} \approx \frac{1}{7.915t}
\end{equation}
For most experiments, the values of $U$ and $t$ can be more or less independently chosen. While $U$ is primarily fixed by the magnetic field strength and the latter can be chosen such that one is always far from a Feshbach resonance, $t$ can be adjusted by tuning the height of optical lattice. Therefore, the single Bloch band picture remains valid for the purpose of studying the BCS-BEC crossover as depicted by Hamiltonian (\ref{ham}). 

\subsection{Particle-Hole Constraints}
\label{ph_original}
Lattice systems have an additional symmetry stemming from the possibility of describing the physics in terms of either particles or holes; the choice of description is usually made in order to simplify the resulting Hamiltonian. In the case of the Hamiltonian (\ref{hamN}) we can obtain an exact relationship between a system with $n$ fermions (particles) and one with $2-n$ fermions (holes).

Let us for the moment work in the canonical ensemble and look for the ground state of the Hamiltonian (\ref{ham}) with
the constraint that the number of particles per site is $n = n_{\uparrow} + n_{\downarrow}$.  If we now perform the particle-hole transformation 
$c^{\dagger}_{i,\sigma}=(-1)^id_{i,\sigma}$ \cite{definei}, it can be easily verified that the kinetic energy term maintains its form with the replacement of the $c$ operators with $d$ operators.
On the other hand, the on-site interaction term (with the site index omitted for clarity) transforms as
\begin{eqnarray}
-U c^\dagger_{\uparrow}c^\dagger_{\downarrow}c_{\downarrow}c_{\uparrow}\rightarrow
&-&U d^\dagger_{\uparrow}d^\dagger_{\downarrow}d_{\downarrow}d_{\uparrow}\nonumber \\
&+&U \left (d^\dagger_{\uparrow} d_{\uparrow}  + d^\dagger_{\downarrow} d_{\downarrow} \right ) \label{p-h transf}
\end{eqnarray}
Given that $d^\dagger_{\uparrow} d_{\uparrow}  + d^\dagger_{\downarrow} d_{\downarrow} = 2-n$ is fixed in the calculation, the terms in the second line of (\ref{p-h transf}) are constant within the Hilbert space of interest. 
Thus, the Hamitonian maintains its operational form under the particle-hole transformation and the ground state wavefunction for a system of $n$ particles is related to the ground state wavefunction for a system of $2-n$ particles. Their corresponding energies are related as
\begin{equation}
{\cal E}(n)+\frac{Un^2}{4}={\cal E}(2-n)+\frac{U(2-n)^2}{4}
\label{e-ph}
\end{equation}
Differentiating with respect to $n$ we see that the chemical potential, defined as $\mu(n)=\partial{\cal E}(n)/\partial(n)$, satisfies
\begin{equation}
\mu(n)+\frac{Un}{2}=-\mu(2-n)-\frac{U(2-n)}{2}
\label{mu-ph}
\end{equation}

Finally, the thermodynamic potential $\Omega(\mu) = {\cal E}(n(\mu))-\mu\, n$ which is the quantity we shall calculate in the grandcanonical ensemble, satisfies
\begin{eqnarray}
\Omega(\mu)+\mu = \Omega(-\mu-U)-\mu-U
\label{omega-ph}
\end{eqnarray}
We stress that any approximation method used to solve the problem would have to satisfy this symmetry in order to yield physically consistent results. 

\section{Hartree + BCS Theory at $T=0$}
The starting point of the Hartree + BCS theory is the Hamiltonian (\ref{ham}). In mean field theory the contribution of the interaction term $\hat{V}_{\K}=-U 
c^{\dagger}_{\K,\uparrow}c^{\dagger}_{-\K,\downarrow}c_{-\K,\downarrow}c_{\K,\uparrow}$ to the ground state energy can be written as
\begin{widetext}
\begin{eqnarray}
\langle \hat{V}\rangle=-U\sum_{\K}\langle c^{\dagger}_{\K,\uparrow}c_{\K,\uparrow} \rangle\langle c^{\dagger}_{-\K,\downarrow}c_{-\K,\downarrow} \rangle -U\sum_{\K}\langle c^{\dagger}_{\K,\uparrow} c^{\dagger}_{-\K,\downarrow}\rangle\langle c_{-\K,\downarrow}c_{\K,\uparrow} \rangle =-{Un^2\over 4}-U\sum_{\K}F_{\K}F^*_{\K}
\end{eqnarray}
\end{widetext}
The first term is a constant and is the Hartree correction to the ground state energy. Note that since the chemical potential $\mu$ is the derivative of the ground state energy w.r.t. $n$, we can absorb the overall shift of the ground state energy due to the Hartree term in $\mu$ by adding $(nU/2)$ to it. The quantity $F_{\K}$ in the second term is self-consistently obtained by minimizing the ground state energy w.r.t. $F_{\K}$ and this gives: $F_{\K}=\Delta_0/2E_{\K}$. The gap and number equations in HBCS respectively read,
\begin{eqnarray} 
{1 \over U}&=&\sum_{\K}{1\over 2E_{\K}} \label{gaphmft} \\
n&=&\sum_{\K}\left( 1-{\xi_{\K}\over E_{\K}}\right) \label{numhmft}
\end{eqnarray}
where $E_{\K}=\sqrt{\xi_{\K}^2+\Delta_0^2}$. We note that the single particle energies $\xi_{\K}$ in HBCS include a Hartree shift: $\xi_{\K}=\epsilon_{\K}-\mu-nU/2$, where the last term is the Hartree term. It can be easily verified that the HBCS theory satisfies the particle-hole constraints on thermodynamics (see Eqs. \ref{e-ph}, \ref{mu-ph}, and \ref{omega-ph}) derived from the attractive Hubbard model. Eqs. (\ref{gaphmft}) and (\ref{numhmft}) are then self-consistently solved for $\mu$ and $\Delta_0$ for a given value of $n$ \cite{belkhir1, belkhir2}. The result for the chemical potential is plotted in Fig. (\ref{mu_comp}) for a given filling. The chemical potential is monotonically suppressed as a function of coupling. Within the HBCS theory the strong coupling expansion of the chemical potential and the gap are respectively given by: $\mu_{\textup{HBCS}} \simeq -(U/2)+12(t^2/U)(n-1)$ and $\Delta_{\textup{HBCS}} \simeq (U/2)\sqrt{1-(1-n)^2}-(6t^2/U)\sqrt{1-(1-n)^2}$.  We next develop a diagrammatic formulation to include the effects of quantum fluctuations on top of the HBCS theory. 

\section{Diagrammatic method for including quantum fluctuations about HBCS theory}
\label{diagram_section}
\begin{figure}
\includegraphics[width=3in]{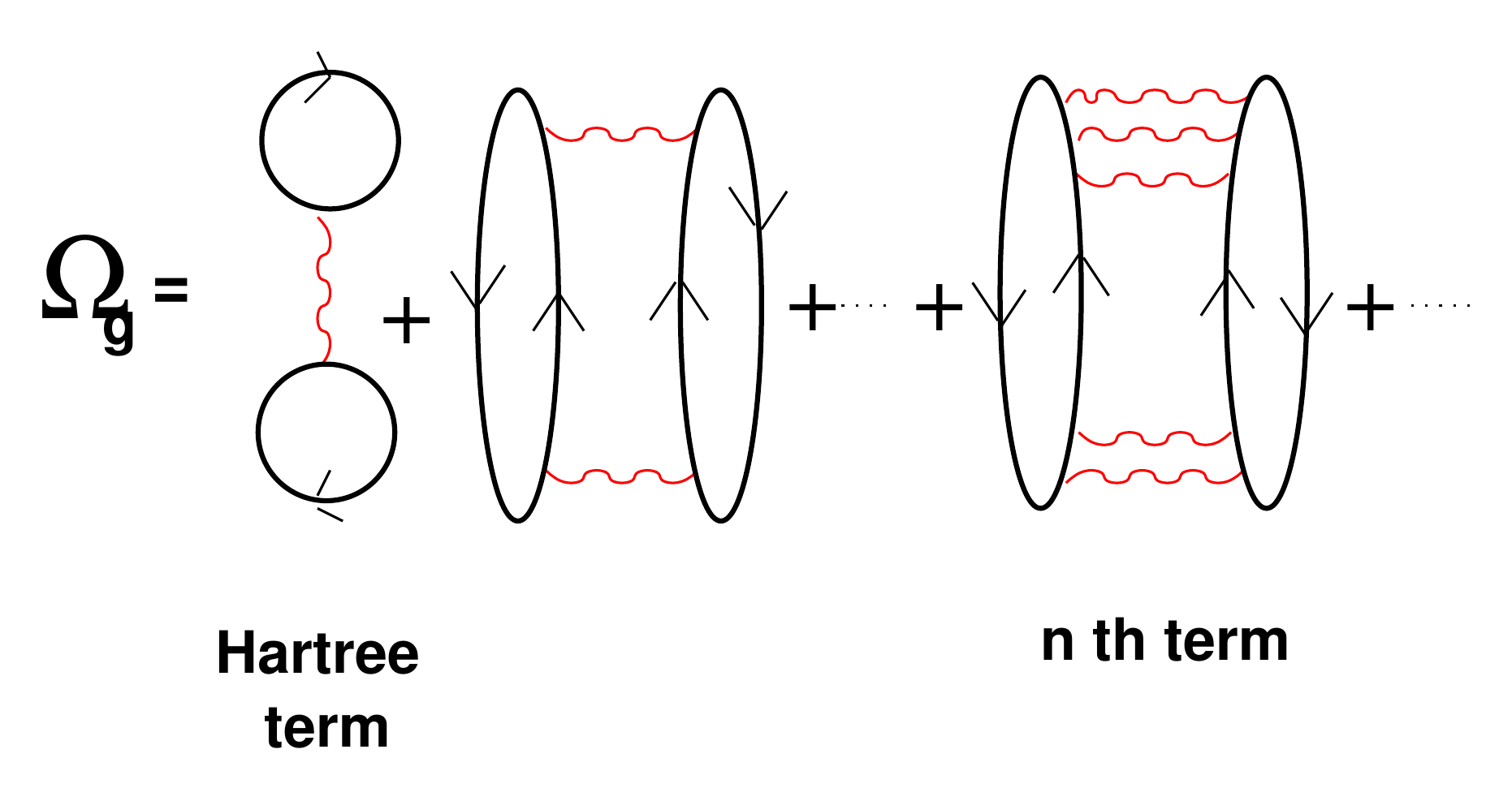}
\caption{RPA loop expansion of the thermodynamic potential in the BCS limit. The first diagram is the Hartree term.}
\label{fig:RPAdiagrams}
\end{figure}

In this section, we outline a diagrammatic approach to include Gaussian fluctuations on top of the HBCS theory. By including quantum fluctuations we expect to account for the zero-point motion of the collective mode and the virtual scattering of gapped quasiparticles. However, since we have already included in our HBCS theory the leading order Hartree term from the Gaussian corrections (see Fig. (\ref{fig:RPAdiagrams})), we should be careful not to double count it. 
We begin by noting that the Hartree term that we have included in the HBCS theory can be systematically introduced in the mean field propagators using the Luttinger-Ward formalism. The details are outlined in Appendix \ref{diagram_appendix}. We next use the Hartree shifted propagators to include RPA corrections to the thermodynamic potential. In order to avoid double counting of the Hartree term, we subtract by hand this term from the Gaussian thermodynamic potential $\Omega_g$. We have explicitly verified that by doing so, we not only restore the correct p-h symmetry for $\Omega_g$, the resulting expression for $\Omega_g$ is also rendered manifestly convergent in the absence of convergence factors which makes it easier to compute $\Omega_g$.   

However, inspite of the compactness of this approach it leads to an unphysical negative compressibility in the strong coupling limit (see Fig. \ref{muvsn}). The result is unphysical since the system in this limit is comprised of hard-core bosons with nearest neighbor repulsion and can hence neither collapse (prevented by Pauli exclusion) nor phase-separate (ruled out on energetic grounds). The failure of the diagrammatic formalism, which was to include quantum fluctuations on top of HBCS theory, therefore necessitates a different approach and we next turn to a large-$N$ formalism.

\section{large-$N$ theory for crossover on lattice}
In this section we give a brief account of the large-$N$ formalism, which as we shall see, starts with a saddle point that is different than HBCS, obeys the p-h constraints appropriate for the large-$N$ model at zeroth and first order in $1/N$, and most importantly predicts positive compressibility for all parameters. In addition to satisfying the p-h constraints on the lattice, the way our large-$N$ theory on the lattice differs from other large-$N$ approaches in the continuum \cite{vsr} is the way we treat the fluctuation feedback (see subsection B). Additionally, at half-filling there is an emergence of charge density wave (CDW) order that the large-$N$ theory is unable to capture (for reasons discussed later). Hence, we shall work away from half-filling where the ground state of the system is a non-degenerate superfluid.  
 
The starting point of our large-$N$ formalism is a generalization of the Hamiltonian of Eq. (\ref{ham}) to include $N$ fermion flavors for each spin in the form
\begin{eqnarray}
H&=&-t\sum_{\langle i,j \rangle, \alpha, \sigma}
(c^{\dagger}_{i\alpha,\sigma}c_{j\alpha,\sigma}+{\textup h}.{\textup c.})\\\nonumber &-&
\frac{U}{N}
\sum_{i,\alpha,\alpha'} c^{\dagger}_{i\alpha,\uparrow}c^{\dagger}_{i\alpha,\downarrow}
c_{i\alpha',\downarrow}c_{i\alpha',\uparrow}-\mu\sum_{i, \alpha,\sigma}n_{i\alpha,\sigma}
\label{hamN}
\end{eqnarray}
where $\alpha$ is the index for each of the $N$ flavors. This Hamiltonian is invariant under the Sp($2N$) symplectic group and reduces to the original attractive Hubbard model (\ref{ham}) after setting $N=1$. As shown in section \ref{func_int}, the virtue of working with the above form of interaction, where the flavor index $\alpha$ is not conserved, is that it lends itself to a systematic expansion in the parameter $1/N$ around the mean field theory results, exact in the limit $N\rightarrow\infty$. Although such an expansion is strictly valid in the large-$N$ limit, it is assumed that the general trends of the results found will be correct after setting $N=1$ at the end of the calculation.

\subsection{Particle-Hole constraints for the large-$N$ model}
\label{ph_N}
Following the discussion in section (\ref{ph_original}), we next derive a set of p-h constraints appropriate for the large-$N$ model (\ref{hamN}). Using the Hamiltonian (\ref{hamN}) and a p-h transformation: $c^{\dagger}_{i\alpha,\sigma}=(-1)^id_{i\alpha,\sigma}$, we obtain exact relationships between thermodynamic variables with $n$ fermions (particles) per flavor and ones with $2-n$ fermions (holes) per flavor. The ground state energy ($\cal E$), the chemical
potential ($\mu$) and the thermodynamic potential ($\Omega$) now respectively transform as follows:  
\begin{eqnarray}
{\cal E}(n)+\frac{Un^2}{4}&=&{\cal E}(2-n)+\frac{U}{4}(2-n)^2 \label{e-ph-N} \\
\mu(n)+\frac{nU}{2N}&=&-\mu(2-n)-\frac{U}{2N}(2-n) \label{mu-ph-N} \\
\Omega(\mu)+\mu N &=& \Omega(-\mu - \frac{U}{N}) +(-\mu -  \frac{U}{N})N
\label{omega-ph-N}
\end{eqnarray}
We next develop a functional integral formalism with the large-$N$ model and show that it respects the above constraints at zeroth order and also at $O(1/N)$.

\subsection{Functional Integral Formalism}
\label{func_int}
In this section we shall outline the key steps in formulating a functional integral approach with the large-$N$ model. The details are given in Appendix \ref{largeN_details}. The thermodynamic properties of the system can be obtained from the partition function which can be expressed as a Feynman path integral over Grassmann fields ${\bar \Psi}_{\alpha\sigma}$ and $\Psi_{\alpha\sigma}$. We next introduce a Hubbard-Stratonovich field $\Delta(x)$ at each $x = ({\bf x}_i, \tau)$ which couples to $\sum_{\alpha} \bar{\Psi}_{i\alpha\uparrow}(\tau)\bar{\Psi}_{i\alpha\downarrow}(\tau)$, and decouple the quartic fermionic interaction term in the action. This makes the functional integral both Gaussian in the fermionic fields and diagonal in the flavor index $\alpha$.  After integrating over these Grassmann variables we get an effective action in terms of the Hubbard-Stratonovich fields $\Delta(x)$. It can be easily shown that the space- and time-independent saddle point of this effective action corresponds to a thermodynamic potential that is linear in $N$ and Gaussian fluctuation corrections to the saddle point are zeroth order in $N$, so that the total thermodynamic potential per flavor can be expanded as
\begin{equation}
{\Omega \over N} = \Omega_{0} + {1\over N} \Omega_g + \cdots 
\label{Omega_N_expansion}
\end{equation}

To find the uniform, static saddle point of the effective action $S_\Delta$, we replace $\Delta(x)$ by the space-time independent quantity
$\Delta_0$. The saddle point condition is~\cite{saddle point comment} $d S_0 / d \Delta_0 = 0$, which can be rewritten as
\begin{equation}
\frac{1}{U}=\sum_{\bf k}\frac{1}{2E_{\bf k}} \label{gapN} 
\end{equation}
where $E_{\K}=\sqrt{\xi_{\K}^2+\Delta_0^2}$. The mean field number equation can be obtained from the mean field thermodynamic potential $\Omega_0$ as
\begin{equation}
\left(\frac{\partial \Omega_0}{\partial \mu}\right)_{T, V}=-n \,\,\,\,\,\,\textup{or}\,\,\,\,\,\, n=\sum_{\bf k}\left(1-\frac{\xi_{\bf k}}{E_{\bf k}}\right) \label{numberN}
\end{equation}
Eqs. (\ref{gapN}) and (\ref{numberN}) must be solved self-consistently 
to obtain the mean field gap parameter $\overline{\Delta_0}$ corresponding to the mean field chemical potential $\overline{\mu}$, as well as finding the chemical potential which yields the desired density $n$. The results of this calculation are presented as dashed lines in Fig. \ref{mudelta}. 

It is instructive to show that this mean field theory satisfies the particle-hole constraints in the lattice (see section II(B)) to the proper order, i.e. to zeroth order in $1/N$. From Eq. (\ref{mu-ph}) we see that this corresponds to the chemical potentials on particle and hole sides being related by $\overline{\mu}(n)=-\overline{\mu}(2-n)$.  The validity of this equation can be seen by replacing $\mu \rightarrow -\mu$ without modifying  $\Delta_0$; this leaves (\ref{gapN}) unchanged while replacing $n \rightarrow 2-n$ in (\ref{numberN}).

The large $U/t$ limit of this theory can be easily obtained from the equations. To zeroth order in $t/U$, the chemical potential becomes $\overline{\mu} = (1-n) U/2$ and the gap parameter is $\overline{\Delta_0} = \sqrt{1-(1-n)^2} \, U/2$.
We finally emphasize that the essential difference between the previously described HBCS theory and this mean field theory is the absence of the Hartree term in the latter. Such a term, which corresponds to the particle-hole channel cannot be easily obtained at the mean field level of any functional integral formalism. We recover this important contribution in our theory as a $1/N$ order correction in what follows.

We next expand our action in terms of the fluctuations around the saddle point and truncate to Gaussian order or $O(1/N)$. The Gaussian thermodynamic potential $\Omega_g$ can be expressed in terms of the fluctuation propagator (see Appendix \ref{largeN_details}), which has poles on the real axis corresponding to the collective mode and branch cuts corresponding to the two-particle continuum. Using the new approximation to the thermodynamic potential per flavor, $\Omega_0(\mu) + (1/N) \,\Omega_g(\mu)$, we next obtain expressions for the properties of the system to linear order in $1/N$. At this point, we want to emphasize that we do not treat the chemical potential $\mu$ and the auxiliary field $\Delta_0$ at equal footing~\cite{feedback}. Indeed, in our approach the former is a thermodynamical variable while the latter is merely a parameter in the theory, obtained as the saddle point of a variable that is integrated over in the partition function. As such, it is not an independent variable but it is defined as a function of $\mu$, i.e. $\Delta_0(\mu)$ is the saddle point field used to calculate the partition function at such a chemical potential, obtained from the solution of Eq. (\ref{gap}). As we make expansions in powers of $1/N$ this equation is left unchanged, as the saddle point condition is exact to all orders~\cite{saddle point comment}.

In order to calculate the leading order corrections to the thermodynamic quantities we next expand the renormalized number equation and the saddle point condition using: $\mu=\mu_0+(1/N)\delta\mu$ and $\Delta=\Delta_0+(1/N)\delta\Delta$. This gives us the Gaussian corrections $\delta\mu$ and $\delta\Delta$ to the chemical potential and gap parameter respectively. In order to make a connection to the original system with two spin components, we set $N=1$.

Finally, we note that, even though our approach to the $1/N$ expansion is different from the one introduced in reference \cite{vsr} it can be shown (see Appendix \ref{vsr_appendix}) that the first order corrections to the chemical potential, $\delta\mu$, in both approaches are equivalent while the corrections to the gap parameter $\delta\Delta_0$ are different; the latter is due to a modification of the gap equation at the $1/N$ level which we do not include. As long as the emphasis is in the calculation of the  thermodynamics of the system which depend solely on the chemical potential, this difference is not as relevant.

\subsection{Zero-temperature Results for the large-$N$ formalism}\label{Results section}
Using our formalism we can calculate all thermodynamic quantities for the system. In this section we present our results, both for the mean field approximation and up to linear order in the $1/N$ expansion; in the figures we have set the number of flavors $N$ equal to 1 in the expansion. 

\subsubsection{Chemical potential and gap parameter}

\begin{figure}[tbp]
\includegraphics[width=3in]{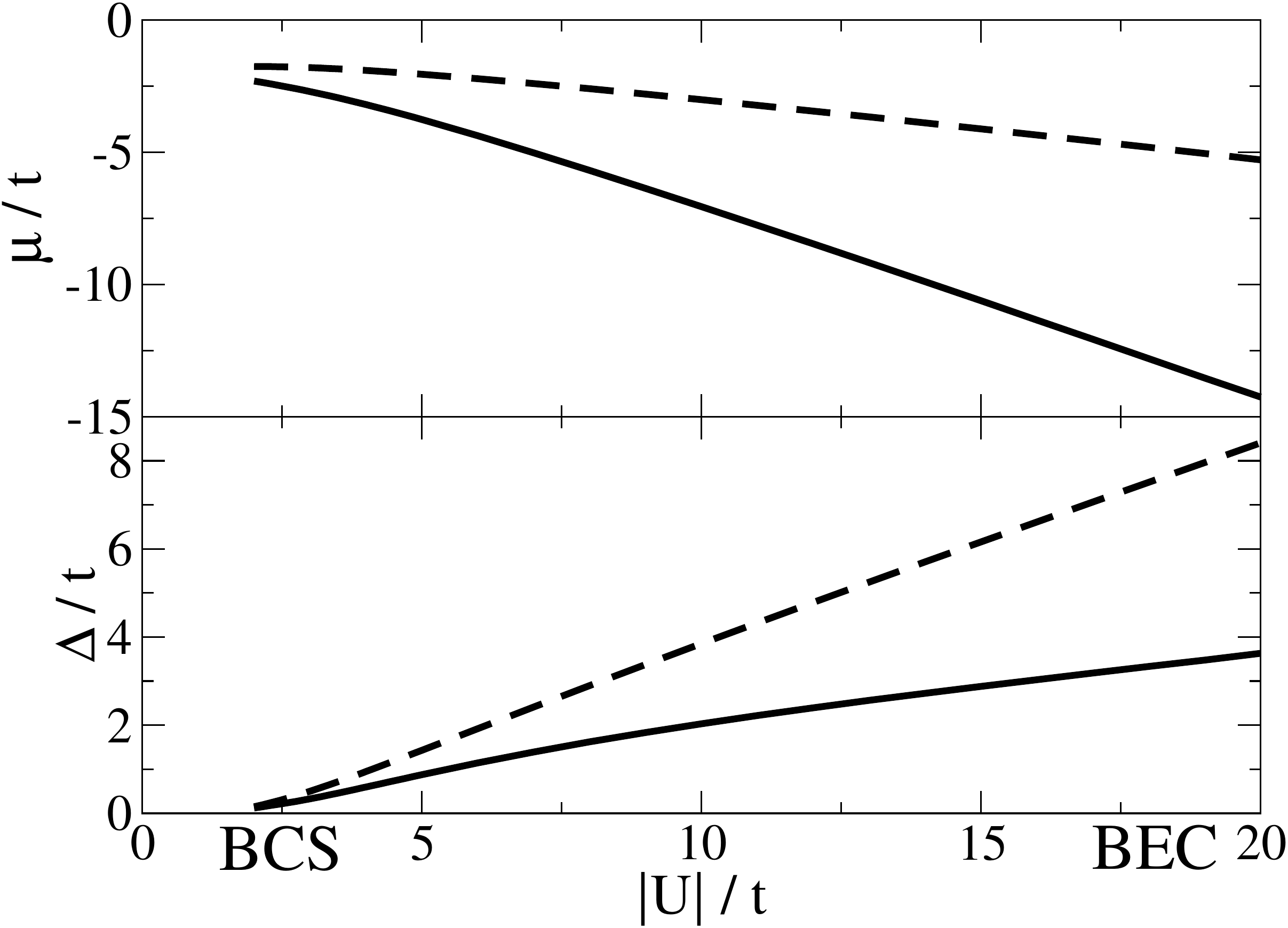}
\caption{Gap $\Delta$ and chemical potential $\mu$ as a function of $U/t$ for a filling $n=0.5$. The dashed line is the mean field result and the solid line is the result which includes fluctuations upto order $1/N$.}
\label{mudelta}
\end{figure}
The chemical potential and the gap parameter across the entire crossover and for a typical density (quarter filling, $n=0.5$) are plotted in Fig. \ref{mudelta}; while the fluctuations are unimportant for small interactions $U$, the correction becomes important at unitarity and in the BEC limit. The fluctuations decrease the value of the order parameter, as well as decrease the value of the chemical potential; as we shall see this is related to the Hartree shift in the energy of the system.

We can show that our theory satisfies particle-hole symmetry {\it to first order in} $1/N$.  As we can see from (\ref{omega-ph-N}) and the expansion (\ref{Omega_N_expansion}), particle-hole symmetry at this order implies that 
\begin{equation}\label{Omega_g_symmetry}
\Omega_g(-\mu) = \Omega_g(\mu) - U (1-n)
\end{equation}
where $\mu = \overline{\mu}(n)$. This property of our $\Omega_g(\mu)$ can be directly seen from the second line in (\ref{omegag}); once again making the transformation $\mu \rightarrow -\mu$ as well as switching variables to ${\bf k} \rightarrow -{\bf k}$ and $\Q \rightarrow -\Q$ we can see the first term is left unchanged while in the second one $u \leftrightarrow v$. Thus, we recover (\ref{Omega_g_symmetry}).

\subsubsection{Compressibility}
\begin{figure}[tbp]
\includegraphics[width=3in]{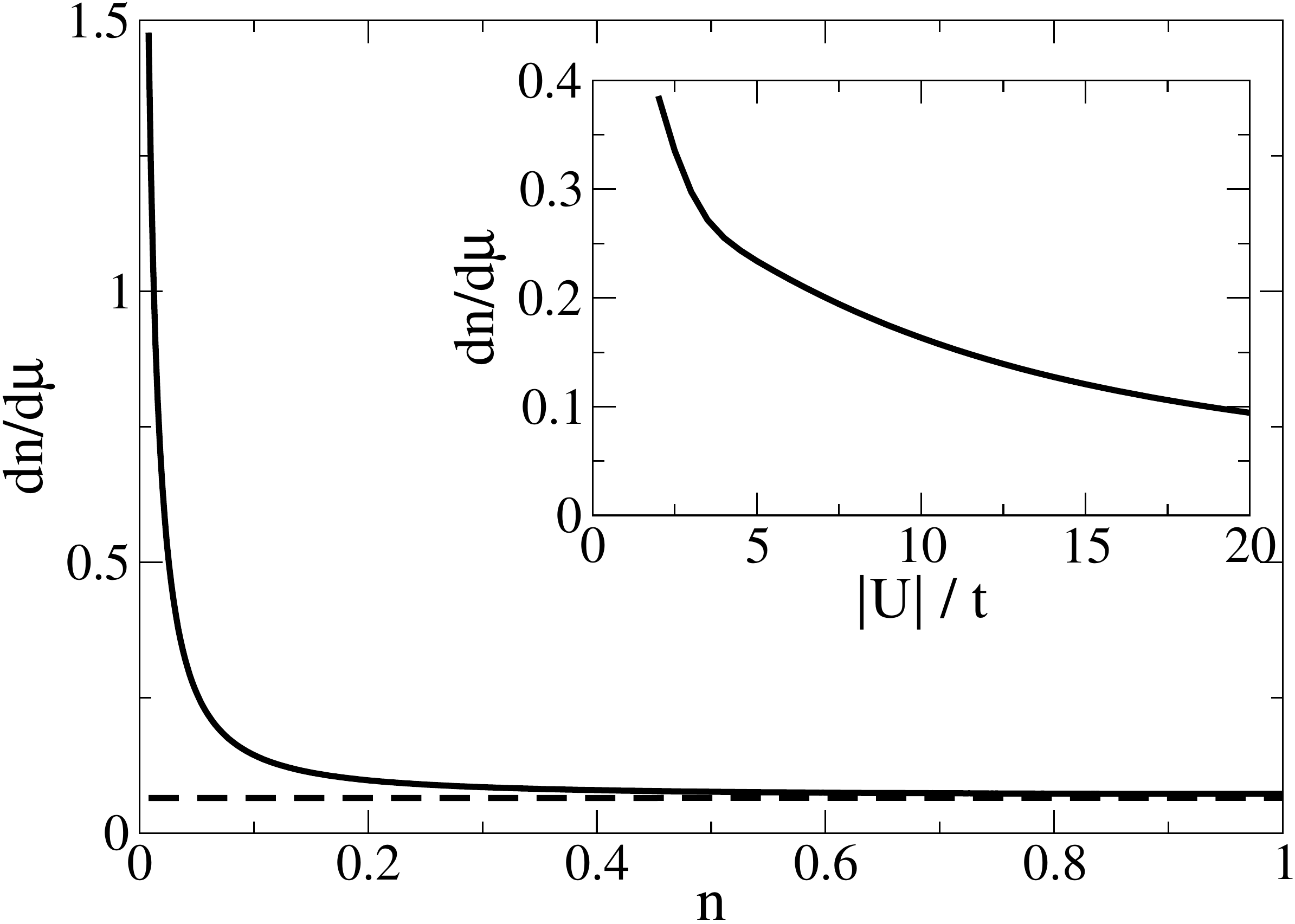}
\caption{The compressibility $dn/d\mu$ as a function of filling $n$ in the strong coupling limit ($U/t=30.0$). The dashed line is the mean field compressibility. The solid line is the compressibility upto $O(1/N)$ and can be seen to be greater than zero and large (but finite) for smaller values of $n$. The inset shows how compressibility changes across the crossover for $n=0.5$.}
\label{cmprssblty}
\end{figure}

We next calculate the compressibility of the system defined as $\kappa=dn/d\mu$ to order $1/N$. 
Using the number equation $n = -d (\Omega_0 + (1/N) \Omega_g )/ d \mu$, differentiating with respect to $\mu$ and evaluating the resulting expression at $\mu = \overline{\mu} + \delta\mu/N$, we obtain
\begin{equation}
\hspace{-.085cm}\frac{dn}{d\mu}=-\frac{d^2\Omega_0}{d\mu^2}-\frac{1}{N}\left[-\left(\frac{d^3\Omega_0}{d\mu^3}\right)\frac{d\Omega_g/d\mu}{d^2\Omega_0/d\mu^2}+\frac{d^2\Omega_g}{d\mu^2} \right]
\end{equation}
Fig. (\ref{cmprssblty}) shows the compressibility as a function of filling in the strong coupling limit $(U/t=30.0)$ both at the mean field level and after including $1/N$ corrections. The positive compressibility in the BEC limit is a check that our theory gives physically correct results in this regime. The inset in Fig. (\ref{cmprssblty}) shows the behavior of compressibility across the crossover. In the strong coupling limit $dn/d\mu \propto 1/U$ stemming from the fact that $\overline{\mu} = (1-n) U/2$ and that fluctuations are unimportant for this quantity.  

\subsection{Comparison between HBCS and large-$N$ results at $T=0$}
\label{hmft_largeN_comp}
\begin{figure}[tbp]
\begin{center}
\includegraphics[width=3in]{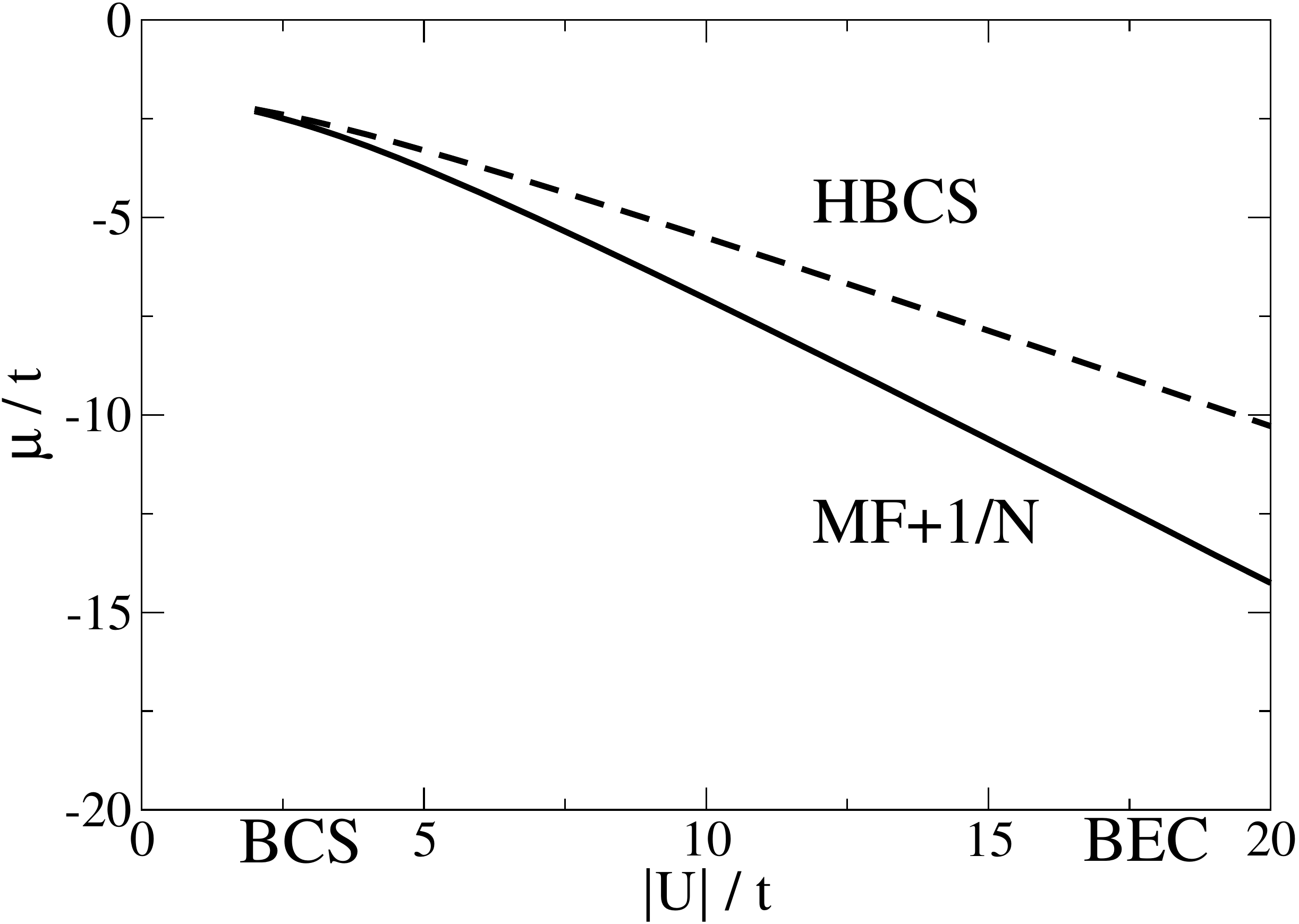}
\caption{Figure shows a comparison of the chemical potentials obtained within Hartree shifted BCS theory and large-$N$ respectively. The filling fraction $n=0.5$.}
\label{mu_comp}
\end{center}
\end{figure}

In this section we shall compare the $T=0$ results from the large-$N$ theory with the HBCS results. To begin with, we plot the chemical potentials for the two theories in Fig. (\ref{mu_comp}). We note that the results from the two theories match in the BCS regime; however there is a large deviation in the BEC regime.  
In particular, when $U/t \gg 1$,
\begin{equation} 
\mu_{\textup{HBCS}} \simeq -(U/2)+12(t^2/U)(n-1)
\label{mu_hbcs}
\end{equation}
On the other hand, the strong coupling limit of the chemical potential within the large-$N$ theory scales like 
\begin{equation}
\mu \simeq -(U/2)(2-n)+O(t^2/U),
\label{mu_largeN} 
\end{equation}
which has a different leading order term compared to the HBCS $\mu$. We already see that there are quantitative differences between the HBCS and the large-$N$ results. 

In order to understand the results (\ref{mu_hbcs}) and (\ref{mu_largeN}), we next solve the problem exactly in the atomic limit: $t/U=0$ and show that the Hartree Mean Field theory gives exact
answers in this limit. For $t/U=0$ the Hamiltonian is a single site one
\begin{equation}
H=-Un_{\uparrow}n_{\downarrow}-\mu (n_{\uparrow}+n_{\downarrow})
\end{equation}
and has four eigenstates: $\mid0\rangle, \mid \uparrow\rangle, \mid \downarrow\rangle$ and $\mid \uparrow\downarrow\rangle$ with
respective energies $0, -\mu, -\mu$ and $-2\mu-U$. To study the broken symmetry state we now introduce a fictitious
pairing field $h$ to obtain a Hamiltonian
\begin{equation}
H=-Un_{\uparrow}n_{\downarrow}-\mu (n_{\uparrow}+n_{\downarrow})-h(c^{\dagger}_{\uparrow}c^{\dagger}_{\downarrow}+c.c.)
\label{newH}
\end{equation}
After setting up the gap and number, we finally take $h=0$, solve the gap and number equations for $\mu$ and $\Delta_0$, and obtain $\mu=-U/2$ and
$\Delta_0=U\sqrt{n(2-n)}/2$. These are exactly the values obtained from solving the HBCS number and gap equations (\ref{gaphmft}, \ref{numhmft}) \cite{garg_dmft}. Physically, the result $\mu=-U/2$ in the atomic limit can be explained by noting that the chemical potential of the fermions is just one-half of the binding energy $\sim U$ for the molecules. From these considerations, we conclude that the large-$N$ theory gives
quantitatively incorrect results for the density ($n$) dependence of the
chemical potential in the atomic limit. This also
leads to problems with the compressibility since the compressibility is the derivative of $\mu$ with respect to $n$. We turn to this next.

In order to calculate the compressibility $dn/d\mu$ within HBCS we note that the Hartree shift to the single particle energies can be incorporated into the chemical potential. Following \cite{trivedi95} we write the renormalized chemical potential as $\tilde{\mu}=\mu+nU/2$, and evaluate $dn/d\mu$ as
\begin{equation}
{dn\over d\mu}={\partial n/\partial{\tilde \mu}\over 1-(U/2)\partial n/\partial{\tilde \mu} }
\label{dndmu1}
\end{equation} 
\begin{figure}[tbp]
\begin{center}$
\begin{array}{ccc}
\includegraphics[width=1.5in]{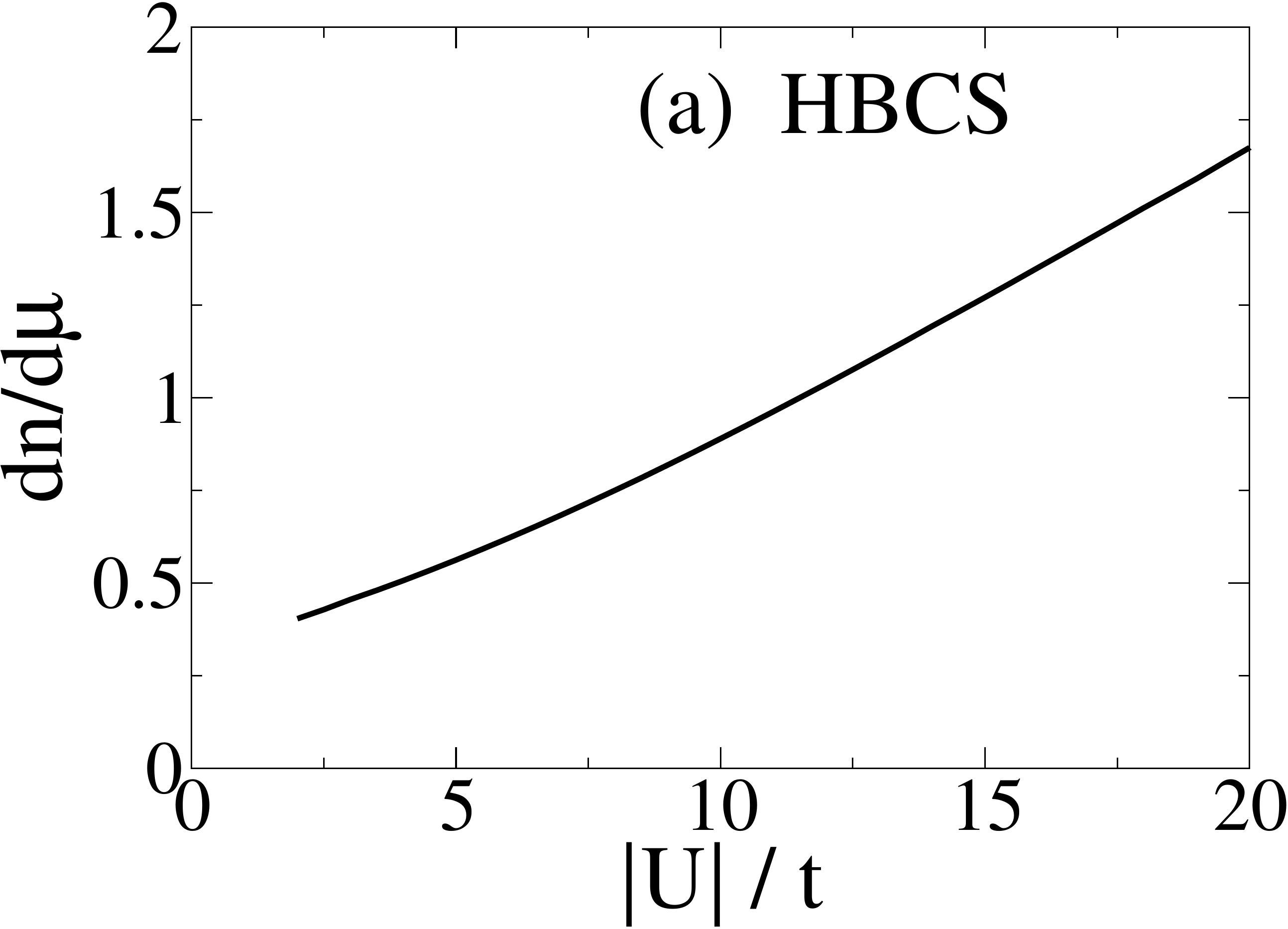}\label{com_comp}&\,\,\,\,&
\includegraphics[width=1.5in]{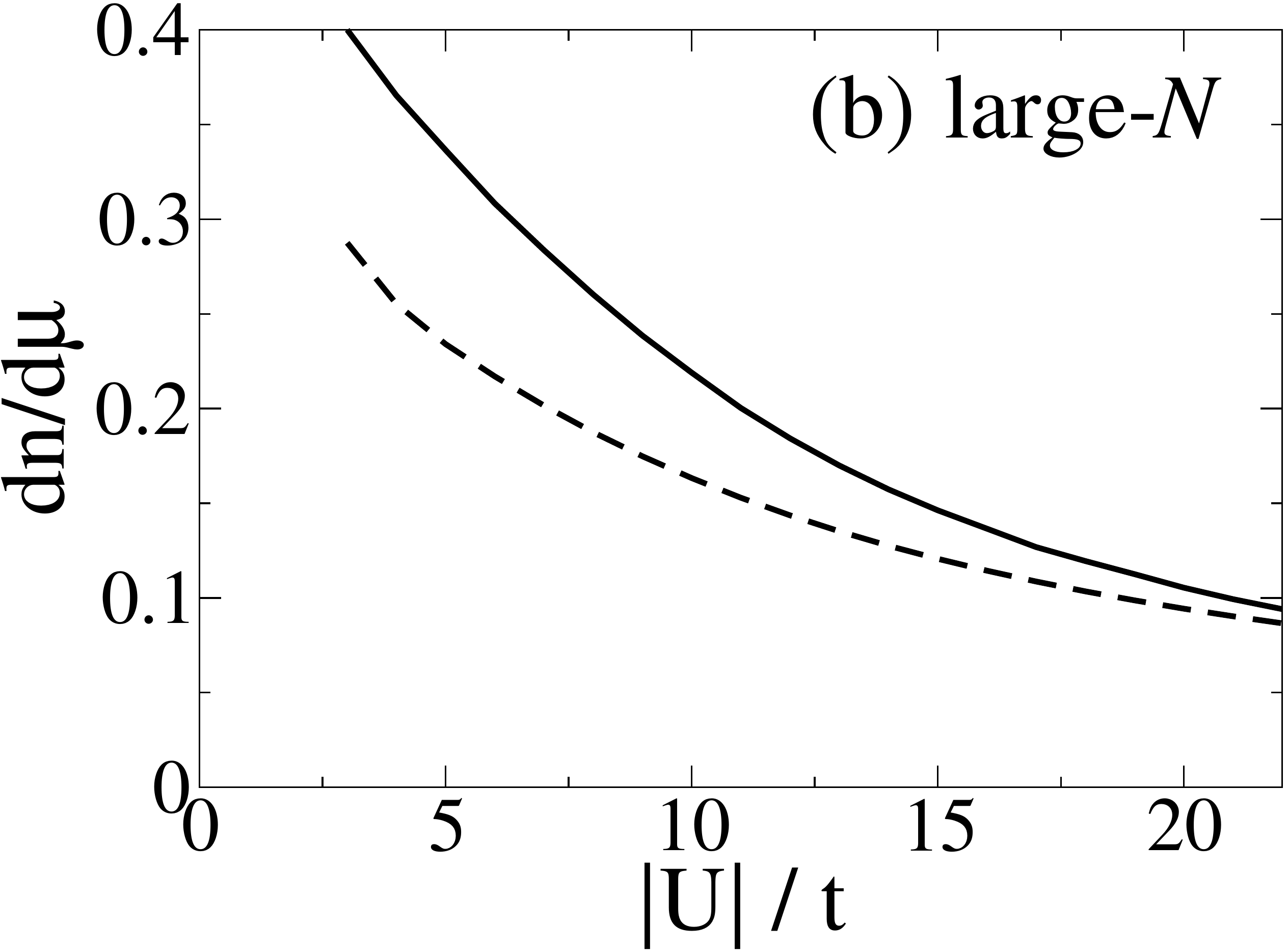} 
\end{array}$
\end{center}
\caption{(a) Compressibility $dn/d\mu$ obtained within Hartree shifted BCS theory for filling fraction $n=0.5$. (b) Compressibility $dn/d\mu$ obtained by using large-$N$ theory for filling fraction $n=0.5$. The dashed line is the MF and the solid line is the MF+$1/N$ compressibility.}
\end{figure}


The quantity $\partial n/\partial {\tilde \mu}$ can be calculated from Eqs. (\ref{gaphmft} and \ref{numhmft})
\begin{equation}
{\partial n \over \partial {\tilde \mu}}=\Delta_0^2\sum_{\K}{1 \over E_{\K}^3} + \frac{(\sum_{\K}\xi_{\K}/E_{\K}^3)^2}{\sum_{\K}1/E_{\K}^3}
\label{dndmu2}
\end{equation}
Fig. (5) shows a comparison of $dn/d\mu$ as obtained from HBCS using Eqs. (\ref{dndmu1}, \ref{dndmu2}) and large-$N$ respectively. In the strong coupling limit the HBCS compressibility scales as
\begin{equation}
{dn\over d\mu}\simeq \frac{U}{12t^2}\left( 1-{24t^2\over U^2}\right)
\end{equation} 
This is again expected on general grounds, when one notes that the chemical potential $\mu$, written in powers of $t/U$, has a zeroth term equal to $-U/2$ (atomic limit). Any $n$ dependence of $\mu$ is hence at least $O(t^2/U)$, which implies that the compressibility must increase with $U$. From these considerations we come to the conclusion that the compressibility should be a monotonically increasing function of $U/t$.
In contrast, the compressibility for large-$N$ scales like $dn/d\mu \sim 2/U$.

To summarize we find that the much simpler BCS plus Hartree theory
works better in the
BEC limit compared with the more sophisticated large-$N$ approach where
we included the $1/N$ Gaussian fluctuation corrections to the saddle
point result, with both approximations satisfying the particle-hole
constraints on the thermodynamics.
By better we mean that BCS + Hartree reproduces the atomic limit
behavior of the chemical potential and the strong coupling behavior of
the compressibility expected for a BEC of hard-core lattice bosons,
while the large-$N$ approach (with $N$ set equal to 1 at the end) does
not. We can also compare our results with the available Quantum Monte
Carlo data, which however is only for the two-dimensional attractive
Hubbard model. We find that the results for the chemical potential
\cite{randeria92} and for the compressibility \cite{trivedi95} at moderately large $\mid U\mid /t$ are both in good
(semi-quantitative) agreement with the Hartree + BCS theory. Although there exists no QMC data on the 3D attractive Hubbard model at $T=0$, we note that fluctuations should be less important in 3D than in
2D. It is therefore reasonable to expect that the agreement between Hartree + BCS theory and QMC should only improve in 3D.

The comparison between the two approaches (large-$N$ and HBCS) is quite surprising and
unexpected. We should emphasize that the two theories start with quite
different mean field solutions (or saddle points). The BCS + Hartree
solution incorporates both the particle-particle (p-p), or pairing,
and the particle-hole (p-h) Hartree physics on an equal footing at the
mean field level. The large-$N$ solution, on the other hand, is designed
to focus only on the p-p channel at the saddle point level, and
include all other effects as fluctuations about the saddle point. One
might have thought that since the Hartree correction to the
thermodynamics is included at the $1/N$ level (along with higher order
terms), the large-$N$ approach would ``go beyond" the simpler BCS +
Hartree approach. But we find that ``more" (diagrams, for instance) is not
necessarily ``better" for quantum many-body systems!

It might also be worth contrasting the optical lattice calculations
presented here from BCS-BEC crossover in the continuum. In the
continuum, one does not in general have a Hartree term in
thermodynamics, which is proportional to both the interaction and the
density (except in the BCS limit). The reason is as follows: the bare
interaction $g(\Lambda)$ actually goes to zero as the ultraviolet
cutoff (inverse range of potential) goes to infinity. Thus a ``bare
Hartree term" proportional to $g(\Lambda) n$ vanishes throughout the
crossover. Also there can be no term proportional to the renormalized
interaction $a_s$ in the ground state energy in general, since that
would diverge at unitarity! As shown in Ref.~\cite{rsr}, the
Hartree diagram in the Gaussian fluctuation correction to the BCS
theory does indeed lead to the expected Hartree correction of relative
order $k_f a_s$ in the BCS limit. But it is not possible to isolate
``the Hartree correction" to the ground state energy or chemical
potential at arbitrary
values of $1/k_f a_s$ in the continuum problem.

In the following section we compute the superfluid density on a lattice at $T=0$ and show that because of the broken translational invariance, the superfluid density is not equal to the total density as is the case with the continuum. 

\section{Superfluid Density}
\label{SFdensity}
It has been shown in Ref.~\citep{LeggettSF}, on quite general grounds, that the superfluid density of a translationally invariant superfluid possessing time reversal invariance at $T=0$, is equal to the total density. For a one component system, barring pathologies (e.g. He$^3$-He$^4$ mixtures), the statement can be proved using the Gibbs-Duhem relation and Landau's two-fluid model. A phase
twist put in the boundary conditions for the order parameter in a
translationally invariant system is uniformly distributed across the system. The
situation is different on a real optical lattice - because of the broken translation symmetry, the many-body wavefunction has a
very small amplitude between the lattice sites and it is energetically advantageous to distribute the phase twists at these locations. As a result the superfluid density, which is the response of the system to this phase twist,
turns out to be different on a lattice. Indeed one can show, using Kubo formalism, that for a translationally invariant system the paramagnetic part of the current-current
correlation function vanishes due to the commutativity of the total momentum operator with the Hamiltonian. The superfluid density in such a system is therefore
entirely given by the diamagnetic part of the response and turns out to be equal to the
total density. In a discrete lattice model, the total momentum operator does not commute with the Hamiltonian and
hence the paramagnetic part of the response is non-zero. Consequently, the superfluid density differs from the total density on a lattice even at $T=0$.

The superfluid density is computed by comparing the free energy $F(n) = \Omega + \mu n$ of the gas at rest with the free energy of a gas moving with a superfluid velocity ${\bf v}_s = {\bf Q}/ (2m)$ in the limit ${\bf Q} \rightarrow 0$; indeed, $F({\bf Q}, n) - F({\bf 0},n) = {1\over 2} n_s m v_s^2$ so that~\cite{Taylor}
\begin{equation}
n_s = 4m\, \left ({\partial^2 F \over \partial Q^2}(Q\rightarrow 0)\right )_{n} 
\end{equation}
We can relate this derivative of $F$ with derivatives of $\Omega$ recognizing that $F(Q, n) = 
\Omega(Q, \mu(Q, n)) + \mu(Q, n) n$ and thus
\begin{eqnarray}\nonumber
\left ( {\partial^2 F\over \partial Q^2} \right )_{n} &=&\left ( {\partial^2 \Omega \over \partial Q^2} \right )_{\mu} +\left ( \left ( {\partial\Omega \over \partial \mu} \right )_{Q} + n \right ) \, \left ( {\partial^2 \mu \over \partial Q^2} \right )_{n}\\
&=& \left ( {\partial^2 \Omega \over \partial Q^2} \right )_{\mu} 
\end{eqnarray}
where we have used the number equation at $Q = 0$ in the last line.

Following ~\cite{Taylor} we calculate, within our large-$N$ formalism, the superfluid density on a lattice as a response of the system to a phase twist on the order parameter. We give the details of this calculation in Appendix \ref{rhos_details} and present the results here. 

The mean field superfluid density for the large-$N$ theory is given by
\begin{eqnarray}
n_s^0 &=& {1\over N\beta} \left( {\partial^2 S_0 \over \partial \theta^2} \right )_{\mu, \Delta}\\
&=& \sum_{\K}\left( 1-\frac{\xi_{\K}}{E_{\K}} \right ) \cos(k_xa)
\end{eqnarray}
From this formula we can see that in general we have $n_s^0 <n$. The Galilean invariant result in the dilute, continuum limit is recovered in the limit of small density $n$ and interaction $U/t$, in which the chemical potential $\mu$ is near the bottom of the band. Thus, for the momenta contributing to the sum we have $\cos(k_xa) \approx 1$ and $n_s^0 = n$. 

\begin{figure}[t]
\includegraphics[width=3in]{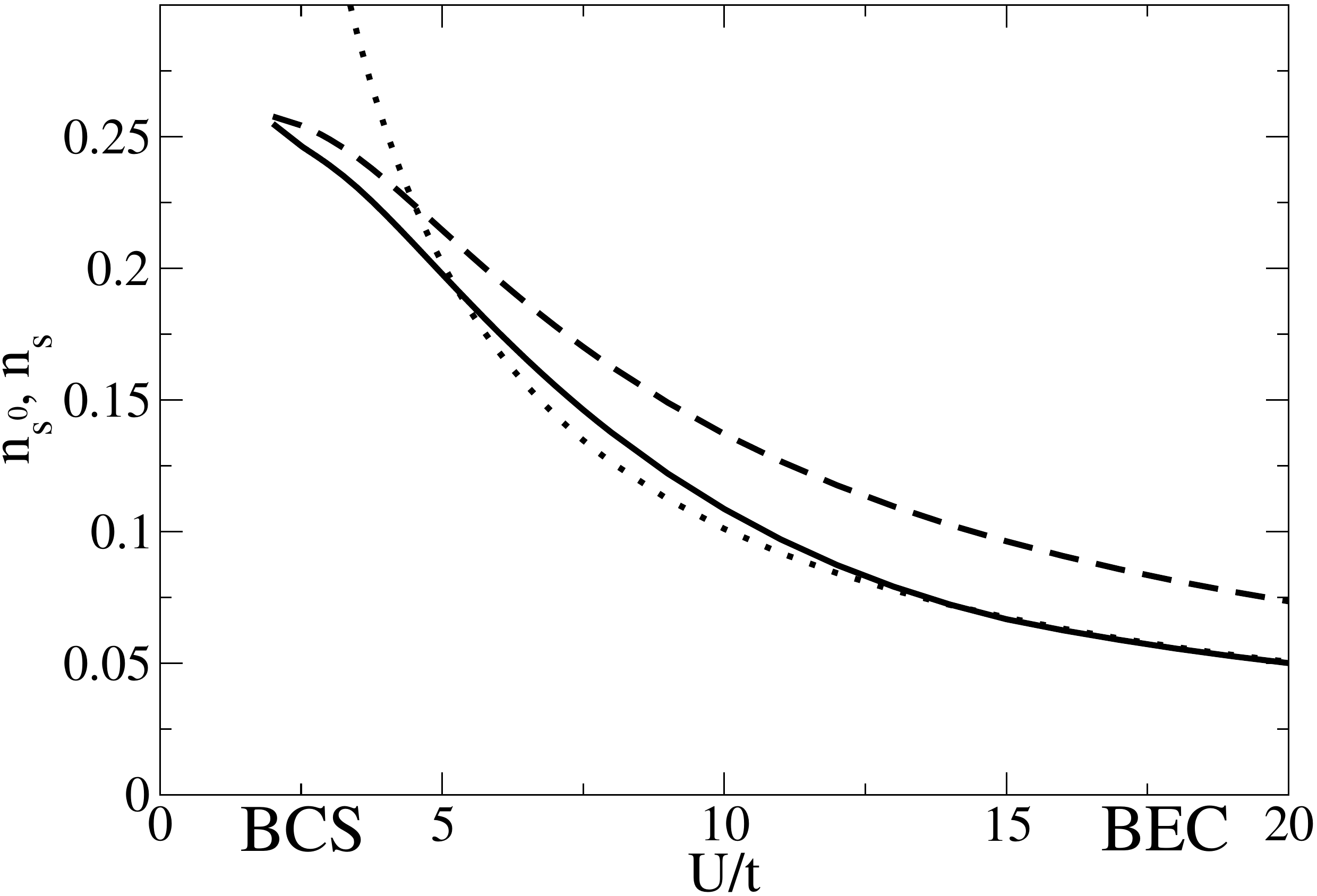}
\caption{The superfluid number density $n_s$ as a function of $U/t$ for $n=0.5$. The dashed line shows the
mean field superfluid number density, the solid line is the result when $1/N$ corrections are included, and the
dotted line is $1.01t/U$ showing that in the strong coupling regime the superfluid density scales like $t^2/U$.}
\label{rhos}
\end{figure}

Next, we obtain the $1/N$ corrections to $\rho_s$ by including Gaussian fluctuations in the calculation of thermodynamic potential (see Appendix \ref{rhos_details} for details). We obtain the following expansion for the superfluid density up to $O(1/N)$:
\begin{widetext}
\begin{eqnarray}
n_s=n_s^0 +\frac{1}{N}\left[ 
{d \over d\mu}\left(\frac{\partial^2{\Omega_0}}{\partial \theta^2}\right)\delta\mu
+\left(\frac{\partial^2{S_g/\beta}}{\partial\theta^2}\right)_{\mu, \Delta_0}+\left(\frac{\partial{S_g/\beta}}{\partial\Delta_0}\right)_{\mu,\theta}
\left( 2 \alpha(\mu) \right)\right] 
\label{omegagtheta}
\end{eqnarray}
\end{widetext}
where $d/d\mu=\partial/\partial\mu+(\partial\mu/\partial\Delta_0)\partial/\partial\Delta_0$ and $\alpha(\mu)$ is given by
\begin{equation}
\alpha(\mu)= -{3\over 8\Delta_0 a^2} \frac{\sum_{\bf k} (\partial^2 \epsilon_{\bf k} / \partial k_x^2)\xi_{\bf k}/E_{\bf k}^3}{\sum_{\bf k} (1/E_{\bf k}^3)}
\end{equation}
We finally set $N=1$ and plot the mean field superfluid density and the one including $1/N$ corrections as a function of coupling strength in Fig. (\ref{rhos}). 
As it can be seen, $n_s$ falls off like $t^2/U$ in the strong coupling limit which can be explained by noting that the system in this regime comprises of hard-core bosons on a lattice with an effective effective mass $\sim U/t^2$. In other words, the tightly bound pairs in the BEC limit can hop only through virtual ionization and hence has an hopping parameter $\sim t^2/U$. Further, Gaussian fluctuations reduce the superfluid number density from its mean
field value across the entire crossover with an increased suppression in the strong coupling regime. This is expected because the BCS mean field theory reduces the problem to one of non-interacting
Bogoliubov quasiparticles with a gapped excitation spectrum. However, the low lying
excitations in the strong coupling regime, are the gapless collective modes of the composite bosons, which are not captured by the BCS mean field theory.

In the next section we calculate the critical temperature within the large-$N$ theory and show that the large-$N$ theory for the attractive Hubbard model, inspite of its above limitations, predicts the correct qualitative trends for $T_c$ in the two limits.

\section{Critical temperature}

Let us finally calculate the critical temperature $T_c$ of the fermionic gas in the lattice, as well as the pairing temperature $T^*$. Just like in the continuum~\cite{sademelo}, these two temperatures are approximately the same in the BCS limit and widely diverging in the BEC limit. In the former, the formation of Cooper pairs and their condensation are governed by the same physics. In the BEC limit, on the other hand, the temperature for the formation of pairs will be of the of the order of the binding energy (which is proportional to $U$) while the critical temperature will be decreasing as $U$ increases, as the effective mass of the pairs increases, as we shall show.


To calculate both $T^*$ and $T_c$ we need to consider the temperature at which the t-matrix has a divergence at zero energy and momentum for a given chemical potential $\mu$. This condition for the inverse temperature $\beta = T^{-1}$ is 
\begin{equation}
\frac{1}{U}=\sum_{\K} \frac{1}{2\xi_{\K}}\tanh(\frac{\beta \xi_{\K}}{2})
\label{tc_gap}
\end{equation}
The two temperatures differ, however, in the equation of state that is used to calculate the density. The pairing temperature is obtain using the mean field approximation to the thermodynamical potential, in which only fermionic excitations are included (i.e. pair breaking at that temperature). The calculation of the critical temperature corresponds to the addition of the effects of Gaussian fluctuations in which bosonic excitations (Goldstone modes) are included, leading to a large renormalization of the equation of state.
In order to set up the number equation we use the same functional integral formalism that we developed at zero temperature.  The mean field thermodynamic potential at high temperatures $\beta < \beta_c$ is:
\begin{equation}
\Omega_0=-\frac{2}{\beta}\sum_{\K} \ln(1+e^{-\beta \xi_{\K}}),
\end{equation}
The mean field number equation at $T=T_c$ is then given by $\partial\Omega_0/\partial\mu=-n$, or 
\begin{equation}
n=\sum_{\K}\frac{2}{\exp(\beta \xi_{\K})+1}.
\label{tc_n}
\end{equation}
Solving Eqs. (\ref{tc_gap},
\ref{tc_n}) self-consistently we obtain the mean field $\beta_c$ for a given density $n$ and denote it by $\beta_c^0 = (T^*)^{-1}$. 

The calculation of the Gaussian contribution to the thermodynamic potential at these high temperatures (at which the gas is normal) is also similar to the one presented at $T=0$ setting $\Delta =0$ everywhere. Thus, the fluctuation propagator $\M$ is diagonal, with $\M_{11}(q)  = \Gamma (q) = 1/U +  \sum_{\K}(1-f-f')/(iq_l-\xi-\xi')$ where $iq_l=i2\pi l/\beta$ are Bose Matsubara frequencies and
$f(\xi)=1/[\exp(\beta \xi)+1]$ is the Fermi distribution function. Hence, 
\begin{equation}
\Omega_g =\frac{1}{\beta}\sum_{\Q, iq_l} \ln\Gamma (q)
\label{tc_omegag}
\end{equation}

Using our large-$N$ formalism, we shall obtain the $1/N$ expansion of the diagrammatic approach to this problem, which was discussed by
Nozieres and Schmitt-Rink (NSR)\cite{nsr}. 
Following NSR, we maintain the same form of the t-matrix equation (\ref{tc_gap}). Nevertheless, just like at $T=0$ the $1/N$ corrections to the thermodynamic potential will renormalize the chemical potential and hence change $\beta_c$. The inverse temperature $\beta$ and the chemical potential $\mu$ are in general independent thermodynamic variables but are related to each other at $T=T_c$ through the t-matrix equation. Expanding Eq. (\ref{tc_gap}) upto $O(1/N)$ we obtain
\begin{equation}
\frac{\partial\Omega_0}{\partial\Delta_0}+\frac{1}{N}\left(\frac{\partial^2\Omega_0}{\partial\Delta_0\partial\mu}\delta\mu+\frac{\partial^2\Omega_0}{\partial\Delta_0\partial\beta}\delta\beta \right)=0
\end{equation}
In the above equation all the derivatives are evaluated at $\Delta=0$, $\mu=\overline{\mu}$ and $\beta=\beta_c^0$. Setting the coefficient of the $1/N$ term to zero we obtain
\begin{equation}
\delta\beta=-\frac{\left(\partial^2\Omega_0/\partial\Delta_0\partial\mu\right)}{\left(\partial^2\Omega_0/\partial\Delta_0\partial\beta\right)}\delta\mu
\label{betacorrection}
\end{equation}
Similarly we expand the number equation and obtain
\begin{equation}
-n=\frac{\partial\Omega_0}{\partial\mu}+\frac{1}{N}\left(\frac{\partial^2\Omega_0}{\partial\mu^2}\delta\mu+\frac{\partial^2\Omega_0}{\partial\mu\partial\beta}\delta\beta+\frac{\partial\Omega_g}{\partial\mu} \right)
\end{equation}
and setting the coefficient of the $1/N$ term to zero we get
\begin{equation}
\delta\mu=-\frac{\left(\partial\Omega_g/\partial\mu\right)+\left(\partial^2\Omega_0/\partial\mu\partial\beta\right)\delta\beta}{\left(\partial^2\Omega_0/\partial\mu^2\right)}
\label{mucorrection}
\end{equation}
Eqs. (\ref{betacorrection}, \ref{mucorrection}) are then simultaneously solved to obtain the $1/N$ corrections to $\beta_c^0$ and $\overline{\mu}$. We obtain the mean field temperature scale $T^*$ and the critical temperature $T_c$ after including $1/N$ corrections from $T^*=1/\beta_c^0$ and $T_c=(\beta_c^0+\delta\beta)^{-1}$ (after setting $N=1$) respectively.
\begin{figure}[tbp]
\includegraphics[width=3in]{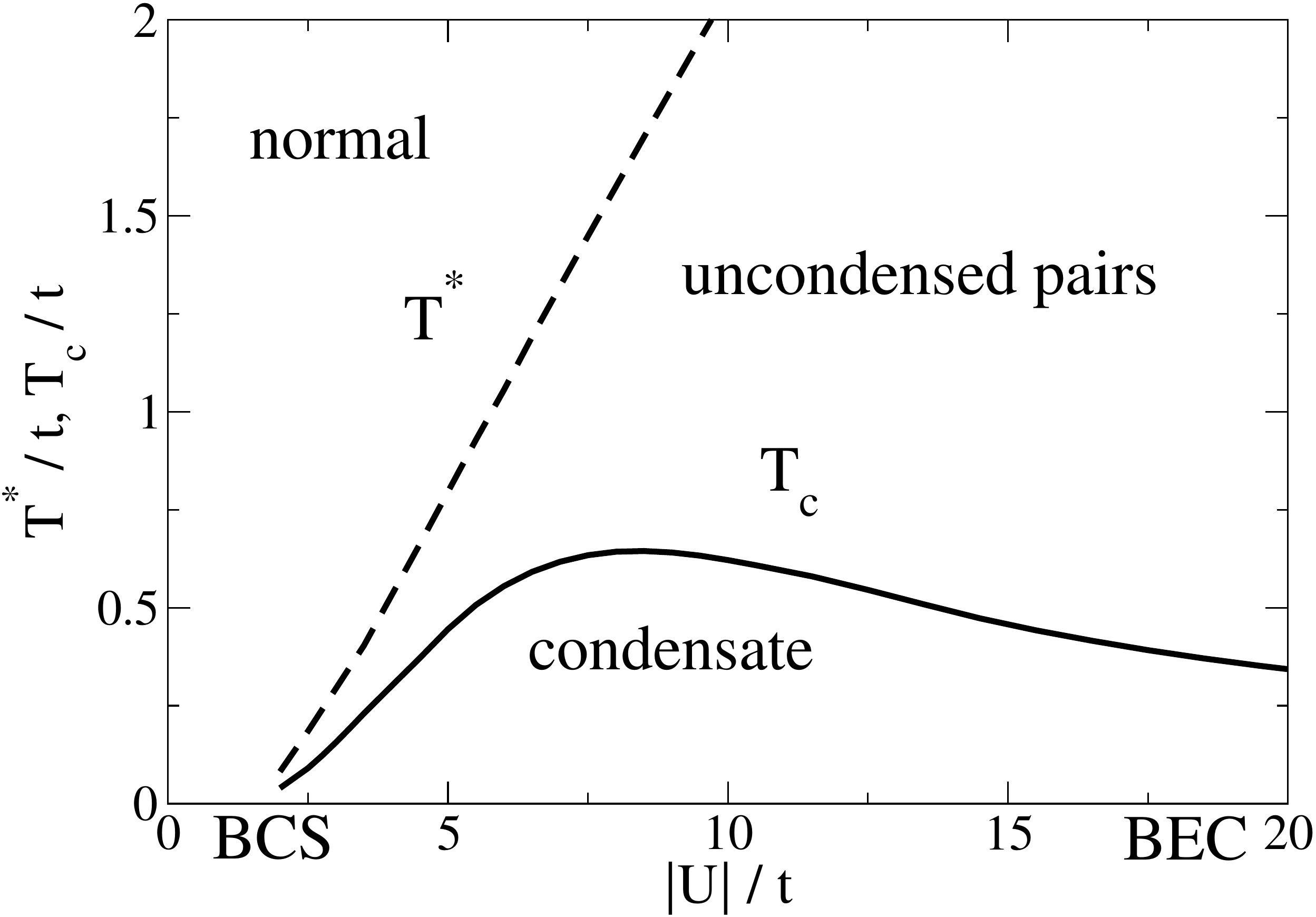}
\caption{The transition temperature $T_c/t$ as a function of $U/t$ for $n=0.5$. The dashed line is the mean field transition temperature denoted by $T^*/t$. The solid line includes fluctuations upto $O(1/N)$.}
\label{tcplot}
\end{figure} 

The results are shown in Fig. (\ref{tcplot}). As expected, there is a large deviation of $T_c$ from its mean field value $T^*$ in the strong coupling regime. The phase diagram is as follows: above $T^*$ the system is a normal fermi gas, for temperatures below $T^*$ and above $T_c$ there are preformed uncondensed pairs and below $T_c$ we have a condensate of pairs. In the weak coupling limit $T_c$ approaches the BCS value $\Delta_0/1.75$. Beyond the BCS regime, the pairing temperature grows linearly with $U/t$ while $T_c$ goes through a maximum near unitarity and then falls off as $t^2/U$. 

\begin{figure}[tbp]
\includegraphics[width=3in]{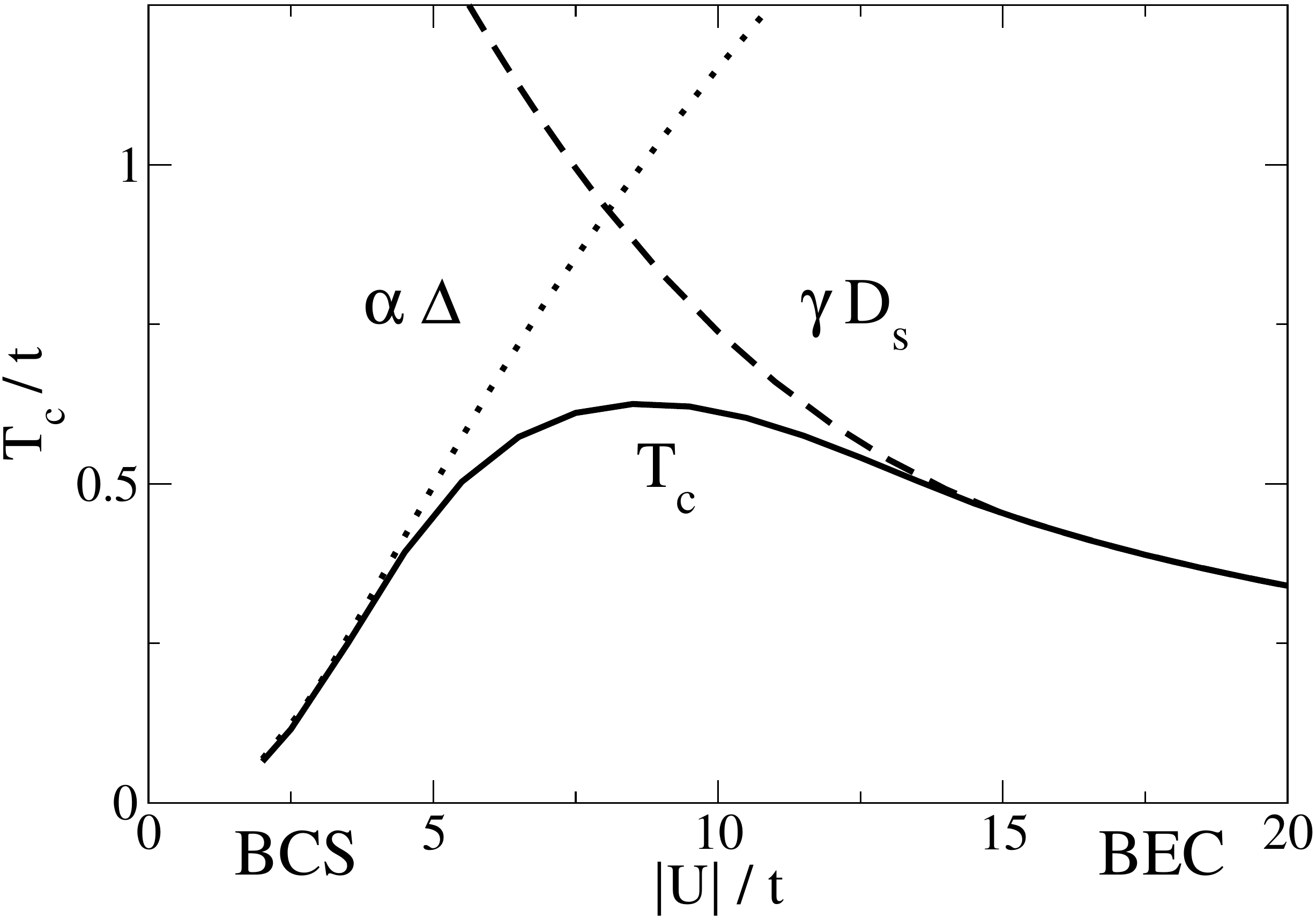}
\caption{This plot shows the scaling of transition temperature $T_c/t$ with coupling $U/t$ ($U$ is on-site interaction and $t$ is hopping matrix element) for a filling of $n=0.5$ in the two limits. The solid black line is
$T_c/t$, the dotted line is the zero temperature gap parameter $\Delta_0/t$ (rescaled by a factor $\alpha=0.57$), the dashed line is zero temperature superfluid stiffness $D_s$ (rescaled by a
factor $\gamma=6.67$). One can see that $T_c$ scales like $\Delta_0$ in the BCS limit and like $n_s$ in the BEC limit.}
\label{scaling}
\end{figure}

\begin{figure}[tbp]
\includegraphics[width=3in]{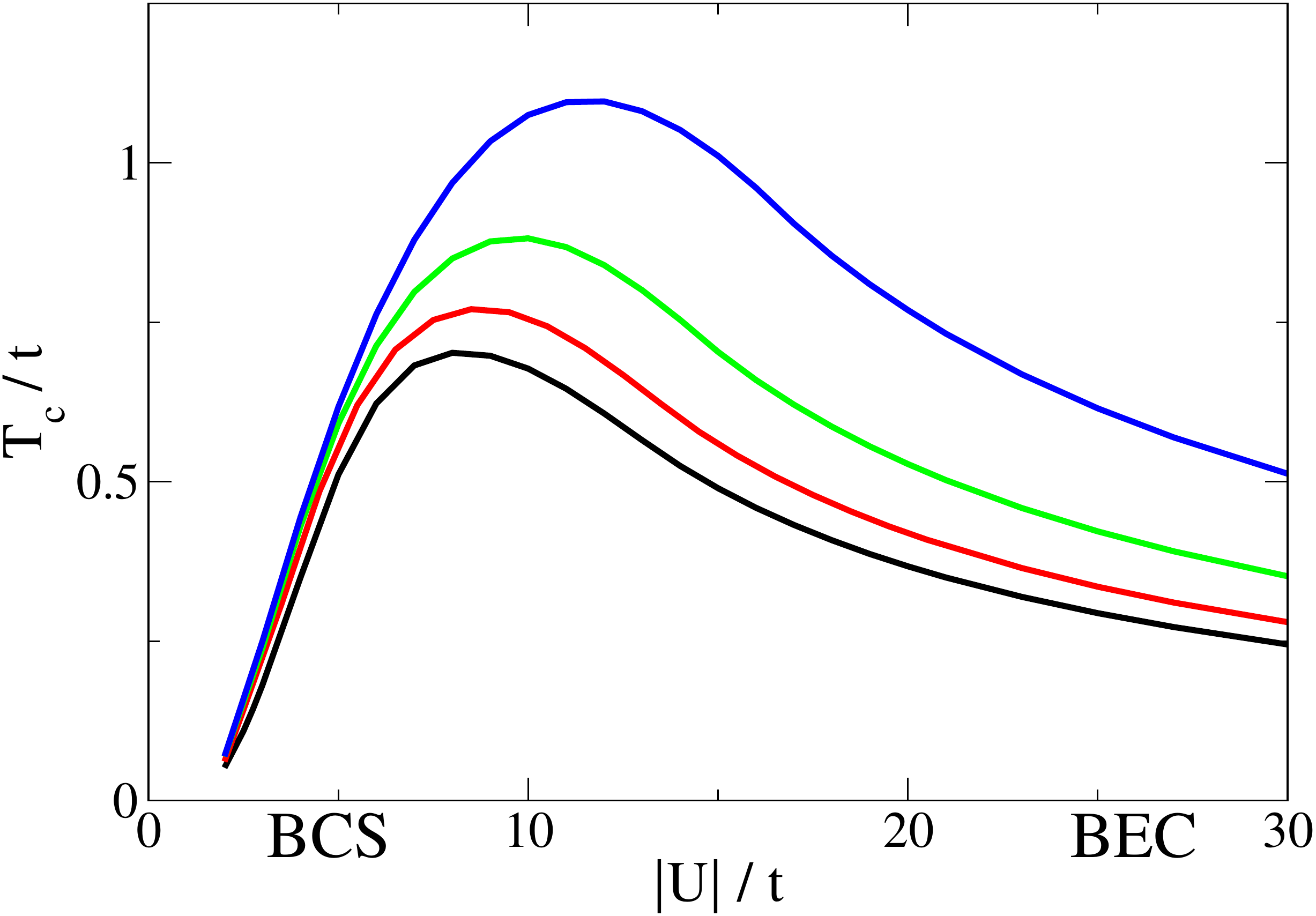}
\caption{(color online) Plot shows $T_c/t$ as a function of $U/t$ for various fillings: $n=0.4$ (black), $n=0.5$ (red), $n=0.6$
(green), $n=0.7$ (blue)}
\label{tccomb}
\end{figure}

The scalings of $T_c$ in the two regimes are summarized in Fig. (\ref{scaling}). In the weak coupling regime the pair breaking energy scale $\Delta_0$ is much smaller than the energy scale set by the superfluid stiffness, $D_s=n_st$ and hence the transition temperature $T_c$ is governed by the zero temperature gap. In the strong coupling regime, the energy scale for phase fluctuations is the smaller one compared to the pair breaking one and hence the scale for $T_c$ is dictated by the zero temperature superfluid stiffness.
The precise value of $U/t$ for which $T_c$ goes to a maximum depends on filling (see Fig. \ref{tccomb}). As explained earlier, the reason why $T_c$ scales like $t^2/U$ in the BEC limit can be explained by considering hardcore bosons on a lattice. A simple second order perturbation theory in $t/U$ then gives an effective hopping parameter proportional to $t^2/U$ for the composite bosons. Since the superfluid stiffness is proportional to the effective hopping parameter and $T_c$ in this regime is governed by phase fluctuations of the lattice Bose gas, therefore $T_c \sim t^2/U$ in this limit~\cite{emery-kivelson, castellani}.
Lastly, we have compared our large-$N$ results with a Hartree shifted NSR theory (HNSR), which uses Hartree shifted propagators in the scattering matrix. The details are given in Appendix \ref{hnsr_appendix}. We find reasonable agreement between the two approaches at BCS and BEC limits (see Fig. \ref{tchnsr}). However, there are quantitative differences between the two results around unitarity which we do not understand at this stage.
\section{Conclusions}
In this paper we have addressed the BCS-BEC crossover on a 3D optical lattice. We have developed a simple Hartree + BCS theory that satisfies the p-h constraints imposed by the lattice on the thermodynamics. Since, inclusion of Gaussian fluctuations on top of the HBCS theory led us to an unphysical negative compressibility in the strong coupling limit, we were forced to start from a different saddle point. We developed a large-$N$ approach for the attractive Hubbard model in 3D, where the large-$N$ saddle point did not include the Hartree shift but still respected the respective p-h constraints of the large-$N$ model. Most importantly, inclusion of Gaussian fluctuations led to a finite and positive compressibility for all parameters. We calculated the ground state chemical potential, gap, ground state energy etc. away from half-filling. The superfluid density at $T=0$ on the lattice was found to deviate from the total density and in the BEC limit was determined by the single-boson hopping matrix element which scales as $t/U$. However, we find that the large-$N$ theory predicts quantitatively inaccurate results for the chemical potential in the strong coupling limit, and qualitatively incorrect trends for the compressibility across the crossover. A comparison with the HBCS theory, which correctly recovers the atomic limit and predicts the right qualitative trends for compressibility, reveals that the large-$N$ theory on the lattice, although considers a larger number of diagrams, is in fact inferior to the simpler Hartree shifted BCS theory. The limitation of the large-$N$ approach is explained by noting (i) the importance of Hartree shift in lattice problems, and (ii) inability of the large-$N$ approach to treat particle-particle and particle-hole channels at equal footing at the saddle point level.  

Inspite of the limitations of the large-$N$ approach in describing the two-component fermionic system on the lattice at $T=0$, we obtain correct trends for the critical temperature within this theory by approaching the superfluid state from above $T_c$. The transition temperature is shown to scale like two different ground state quantities in the two regimes: in the BCS regime $T_c$ scales like the gap, while in the BEC regime $T_c$ scales like the zero temperature superfluid stiffness. These two different scalings show that in the weak coupling regime coherence is lost due to pair breaking and in the strong coupling the superfluid order is destroyed by phase fluctuations of the lattice Bose gas. 
\acknowledgments{The author wishes to thank Roberto Diener, Mohit Randeria, Tony Leggett, Abhishek Mukherjee, and Rajdeep Sensarma for valuable discussions, and the Ohio State University for its hospitality. This work has been supported by NSF through Grant No. NSF-DMR-03-50842.}
\appendix
\section{Diagrammatic Approach for the attractive Hubbard model}
\label{diagram_appendix}
In this appendix we develop a diagrammatic formulation of the crossover problem in the lattice in order to include gaussian fluctuations on top of the Hartree + BCS theory. The starting point of this discussion is the attractive Hubbard Hamiltonian (\ref{ham}). The thermodynamics under this Hamiltonian obeys the p-h constraints discussed in Eqs. (\ref{e-ph}, \ref{mu-ph}, and \ref{omega-ph}). It is easy to see that starting with Hamiltonian (\ref{ham}) if we were to develop a functional integral formalism (like the one used for the large $N$ model), we would find that the saddle point violates the p-h constraints. The reason can be traced back to the choice of Hubbard-Stratonovich field (either p-h or p-p channel). At this stage let us anticipate that a Hartree shift to the mean field chemical potential would correct this problem. We use a Luttinger and Ward formalism \cite{lutward} to systematically introduce a Hartree shift in our single particle propagators. Introducing the Luttinger-Ward functional $\Phi[{\bf G}]$ we write the thermodynamic potential $\Omega$ as
\begin{equation}
\Omega=\Phi+\operatorname{Tr}\ln {\bf G}-\operatorname{Tr} {\bf \Sigma G}
\label{luttinger}
\end{equation}
where $\operatorname{Tr}\equiv (1/\beta)\sum_{{\bf k}, ik_n}\operatorname{tr}$ and the self-energy ${\bf \Sigma}$ is obtained by evaluating the functional derivate of $\Phi[{\bf G}]$ at the exact Green's function
\begin{equation}
\beta\frac{\delta\Phi[{\bf G}]}{\delta{\bf G}}={\bf \Sigma}[{\bf G}]={\bf \Sigma}
\end{equation}
Note that the relation ${\bf \Sigma}[{\bf G}]={\bf \Sigma}$ is independent from the Dyson equation ${\bf G}^{-1}={\bf G}_0^{-1}-{\bf \Sigma}$, where ${\bf G}_0$ is
the non-interacting Green's function. In the Luttinger-Ward formalism $\Phi$ is obtained by summing up an infinite series of closed diagrams without any
self-energy insertion (generally called skeleton diagrams) and replacing all free propagators by fully interacting ones. At the mean
field level we need to retain only the first diagram in the series and thus 
\begin{equation}
\Phi[{\bf G}]=-U(\operatorname{Tr}G_{12})(\operatorname{Tr}G_{21})+U(\operatorname{Tr}G_{11})(\operatorname{Tr}G_{22})
\end{equation}
We define $\delta \Phi[{\bf G}]/\delta G_{11}=U\operatorname{Tr}G_{22}=\Sigma$ which implies $\delta \Phi[{\bf G}]/\delta G_{22}=U\operatorname{Tr}G_{11}=-\Sigma$. We can further
associate $\operatorname{Tr}G_{21}=\operatorname{Tr}G_{12}$ with the Hubbard-Stratanovich field $\Delta$ and therefore $\delta \Phi[{\bf G}]/\delta
G_{12}=-U\operatorname{Tr}G_{21}=-\Delta$. The Luttinger-Ward functional at the mean field level is therefore given by
\begin{equation}
\Phi[{\bf G}]=-\frac{\Delta^2}{U}-\frac{\Sigma^2}{U}
\label{phi} 
\end{equation}
and the self-energy matrix is given by
\begin{equation}
{\bf \Sigma}=\left( \begin{array}{cc}
\Sigma & -\Delta \\
-\Delta & -\Sigma
\end{array} \right)
\end{equation}
We next use the form of the free Green's function given by
\begin{equation}
{\bf G}_0^{-1}=\left( \begin{array}{cc}
ik_n-\epsilon_{\bf k}+\mu & 0 \\
0 & ik_n+\epsilon_{\bf k}-\mu
\end{array} \right)
\end{equation}
and the Dyson equation to calculate the inverse of the full Green's function
\begin{equation}
{\bf G}^{-1}=\left( \begin{array}{cc}
ik_n-\epsilon_{\bf k}+\mu-\Sigma & \Delta \\
\Delta & ik_n+\epsilon_{\bf k}-\mu+\Sigma
\end{array} \right)
\label{fullG}
\end{equation}
Note, the Green's function in Eq. (\ref{fullG}) has its single particle propagators Hartree shifted. 
\subsection{Mean field theory at $T=0$}
Using Eq. (\ref{luttinger}) and the fact that $\operatorname{Tr}{\bf \Sigma}=0$ we can obtain an expression
for the mean field thermodynamic potential 
\begin{equation}
\Omega_0=\frac{\Delta_0^2}{U}-\sum_{\bf k}(E_{\bf k}-\xi_{\bf k})+\frac{\Sigma^2}{U}
\label{omegamf}
\end{equation}
where $E_{\K}=\sqrt{\xi_{\K}^2+\Delta_0^2}$ and $\xi_{\bf k}=\epsilon_{\bf k}-\mu+\Sigma$. This form of thermodynamic potential, as anticipated earlier, obeys the correct
particle-hole constraints. Then the spatially uniform, static saddle
point at $T=0$ is given by the following condition
\begin{equation}
\frac{\partial \Omega_0}{\partial \Delta_0}=0\,\,\,\,\,\, \textup{or} \,\,\,\,\,\, \frac{1}{U}=\sum_{\bf k}\frac{1}{2E_{\bf k}} \label{gap1} 
\end{equation}
The mean field number equation can be obtained from the condition 
\begin{equation}
\frac{\partial \Omega_0}{\partial \mu}=-n \,\,\,\,\,\,\textup{or}\,\,\,\,\,\, n=\sum_{\bf k}\left(1-\frac{\xi_{\bf k}}{E_{\bf k}}\right) \label{number1}
\end{equation}
and the Hartree shift $\Sigma$ is given by
\begin{equation}
\frac{\partial \Omega_0}{\partial \Sigma}=0 \,\,\,\,\,\,\textup{or}\,\,\,\,\,\, \Sigma=-\frac{U}{2}\sum_{\bf k}\left(1-\frac{\xi_{\bf k}}{E_{\bf k}}\right) \label{sigma1}
\end{equation}
Eqs. (\ref{gap1}), (\ref{sigma1}), and (\ref{number1}) are then solved self-consistently and we obtain the mean field values for $\Delta_0$, $\mu$ and $\Sigma$.
\subsection{Gaussian fluctuations at $T=0$}
In order to go beyond the mean field approximation we next consider fluctuations of the order parameter $\Delta$
around its static saddle point value $\Delta_0$
and expand the action $S_{\Delta}$ to Gaussian order. The first order term vanishes due to
the saddle point condition (\ref{gap1}) and we obtain
\begin{equation}
S_{\Delta}=S_0+S_g+...
\end{equation}
The mean field piece $S_0$ has been defined above and Gaussian piece has the form
\begin{equation}
S_g=\frac{1}{2}\sumq (\eta^*(q) \eta(-q))\M(q)\left( \begin{array}{c}
\eta(q) \\
\eta^*(-q)
\end{array} \right)
\end{equation}
where $iq_l=i2\pi l/\beta$ are the Bose-Matsubara frequencies and the matrix elements of the inverse fluctuation
propagator $\M$ are given by
\begin{eqnarray}
\M_{11}(q)=\M_{22}(-q)=1+\sumk \G^0_{22}(k)\G^0_{11}(k+q)  \nonumber \\ 
=\frac{1}{U}+\sum_{\K}\left(\frac{u_{\K}^2u_{\Kp}^2}{iq_l-E_{\K}-E_{\Kp}}-\frac{v_{\K}^2v_{\Kp}^2}{iq_l+E_{\K}+E_{\Kp}} \right) 
\end{eqnarray} and
\begin{eqnarray}\,\,
\M_{12}(q)=\M_{21}(q)=\sumk \G^0_{12}(k)\G^0_{12}(k+q) \nonumber \\ 
=\sum_{\K}u_{\K}u_{\Kp}v_{\K}v_{\Kp}\left(\frac{1}{iq_l+E_{\K}+E_{\Kp}}-\frac{1}{iq_l-E_{\K}-E_{\Kp}} \right)\,\,
\end{eqnarray}
Here $\G^0$ is the same Nambu propagator defined in Eq. (\ref{fullG}) with $\Delta=\Delta_0$, $u_{\K}^2=1-v_{\K}^2=(1/2)(1+\xi_{\K}/E_{\K})$ are the standard BCS coherence factors and $\Kp=\K+\Q$. While calculating the thermodynamic potential including Gaussian fluctuations we need to remember that the first term in the Gaussian part ($\Omega_g$) is indeed the Hartree
term ($-\Sigma^2/U$). Since the Hartree contribution has already been included at the mean field level to preserve particle-hole symmetry, we
need to take it out from $\Omega_g$ to avoid double counting. Writing the partition function upto Gaussian
order and integrating out the Gaussian fluctuations we obtain the Gaussian contribution to the thermodynamic potential 
\begin{equation}
\Omega_g = \frac{1}{2\beta}\sum_{iq_n,{\bf q}}\ln\left( \frac{{\bf M}_{11}}{{\bf M}_{22}}\operatorname{Det}{\bf M}(q)\right)e^{iq_n0+} + \frac{\Sigma^2}{U}
\label{omegag1}
\end{equation}
where the matrix elements $\M_{11}$ etc. have been rescaled as $\M_{11}\rightarrow U\M_{11}$.
It is easy to see that in the limit of large $iq_l$ 
\begin{equation}
\Det \M(q) \sim 1-\frac{U^2}{(iq_l)^2}\left[ \sum_{\K}(u_{\K}^2u_{\Kp}^2-v_{\K}^2v_{\Kp}^2) \right]
\end{equation}
However 
\begin{equation}
\ln \frac{\M_{11}}{\M_{22}} \sim \ln \left[ 1+\frac{2U}{iq_l}\sum_{\K}(u_{\K}^2u_{\Kp}^2-v_{\K}^2v_{\Kp}^2) \right] 
\label{m11m22}
\end{equation}
and hence the Matsubara sum $\sum_{iq_l}\ln \M_{11}\Det\M/\M_{22}$ without the convergence factor diverges for large $iq_l$. However, the correct $\Omega_g$ also
has a correction term given by $\Sigma^2/U$
\begin{eqnarray} 
=-\frac{U}{2\beta}\sumq \frac{1}{\beta}\sumk \left[\G_{11}(\K)\G_{22}(\Kp)+
\G_{11}(\Kp)\G_{22}(\K)\right]  \nonumber \\
=-\frac{1}{2\beta}\sumq \left[ (\M_{22}-1)e^{-iq_l0^+}+(\M_{11}-1)e^{+iq_l0^+}\right] \,\,\,\,\,\,
\end{eqnarray}
Upon changing the sign of $q$ in the second term of the second line and noting that the sum over $q$ is over both positive and negative values we have for large $iq_l$
\begin{equation}
\frac{\Sigma^2}{U}=-\frac{1}{2\beta}\sumq \frac{2U}{iq_l}\sum_{\K}(u_{\K}^2u_{\Kp}^2-v_{\K}^2v_{\Kp}^2)e^{+iq_l0^+}
\end{equation}
which exactly cancels the linear term in the large $(iq_l)$ expansion in Eq. (\ref{m11m22}).

Summing up the above results we obtain the Gaussian correction to the thermodynamic potential
\begin{widetext}
\begin{equation}
\Omega_g=\frac{1}{2\beta}\sum_{iq_n,{\bf q}}\ln\left[\frac{\left( {\bf M}_{11}(q)/{\bf M}_{22}(q)\right)\operatorname{Det}{\bf M}(q)}{\exp\left[
U\sum_{\K}\left\{u_{\K}^2u_{\Kp}^2/(iq_l-E_{\K}-E_{\Kp})-v_{\K}^2v_{\Kp}^2/(iq_l+E_{\K}+E_{\Kp}) \right\} \right]}\right]
\label{finalomegag}
\end{equation}
\end{widetext}
where we are justified to drop the convergence factor $e^{+iq_l0^+}$ from the right hand side of Eq. (\ref{finalomegag}) since in the large $(iq_l)$ limit, the
leading order term in the sum is now of the order $1/(iq_l)^2$ and thus the Matsubara sum is convergent. Thus the
same scheme that restores the correct particle-hole symmetry in our theory, also makes the Matsubara sum convergent
at the Gaussian level. To evaluate the Matsubara sum in Eq. (\ref{finalomegag}) we
analytically continue in the complex plane and convert the sum over the bosonic Matsubara frequencies to an integral over a closed contour enclosing the imaginary axis
counter clockwise $\sum_{iq_l}\rightarrow\oint (dz/2\pi i)n_B(z)$
where $n_B(z)$ is the Bose distribution function. We evaluate the integral over $z$ along a contour parallel to the
Matsubara axis: $z\rightarrow0^-+iy$ keeping in mind that the phase of $\ln \M_{11}(\Q, y)/\M_{22}(\Q, y)$ and the
imaginary part of $(\M_{11}(\Q, y)-1)$ are both odd functions of $y$ and hence do not contribute when integrated
over positive and negative values of $y$. Therefore, we obtain at $T=0$
\begin{eqnarray}
\Omega_g=\frac{1}{2\beta}\sumq \ln\left(\frac{\M_{11}\Det\M}{\M_{22}\exp(-\Sigma^2/U)}\right)\hspace{.5in}\nonumber \\
=\int_0^{\infty}dy/(2\pi)\sum_{\Q}
\left[\ln\left(\Det\M(y)\right)-2\operatorname{Re}(\M_{11}-1)\right]\,\,\,\,\,\,\,\,
\end{eqnarray}
To obtain $\Delta_0$, $\mu$ and $\Sigma$ including gaussian corrections we start with a grand canonical ensemble and treat both $\mu$ and $\Sigma$ as thermodynamic variables. For convinience we switch to $\mut=\mu-\Sigma$ and $\Sigma$ as our independent variables. Then, the thermodynamic potential can be written as $\Omega(\mut, \Delta(\mut); \Sigma)=A(\mut, \Delta(\mut))+\Sigma^2/U$, where the function $A(\mut, \Delta(\mut))$ has no explicit dependence on $\Sigma$. The gap $\Delta(\mut)$ is obtained from the saddle point Eq. (\ref{gap1}). To obtain the number equation and the equation for $\Sigma$ we construct a function $F(\mut, \Sigma)=\Omega(\mut,\Sigma)+(\mut+\Sigma)n$. The condition for $\Sigma$ is then given by
\begin{equation}
\left( \frac{\partial F}{\partial \Sigma} \right)_{\mut}=0 \,\,\,\,\, \textup{or} \,\,\,\,\, \Sigma=-\frac{nU}{2}
\label{sig}
\end{equation}
The number equation reads
\begin{equation}
\left( \frac{\partial F}{\partial \mut} \right)_{\Sigma}=0 \,\,\,\,\, \textup{or} \,\,\,\,\, \left(\frac{\partial A}{\partial\mut}\right)_{\Sigma}+\left(\frac{\partial A}{\partial\Delta}\right)_{\Sigma}\left(\frac{\partial \Delta}{\partial\mut}\right)_{\Sigma}+n=0
\label{mut}
\end{equation}
We next switch to a canonical ensemble and for a fixed $n$ numerically calculate $A[\mut, \Delta(\mut)]=A_0[\mu, \Delta(\mu)]+A_g[\mu, \Delta(\mu)]$. Eq. (\ref{mut}) then gives the value of the renormalized Hartree shifted chemical potential $\mut$ for the corresponding value of $n$ which when combined with Eq. (\ref{sig}) gives the renormalized chemical potential $\mu$ without the Hartree shift.
\begin{figure}[tbp]
\vspace{1.5cm}
\includegraphics[width=3in]{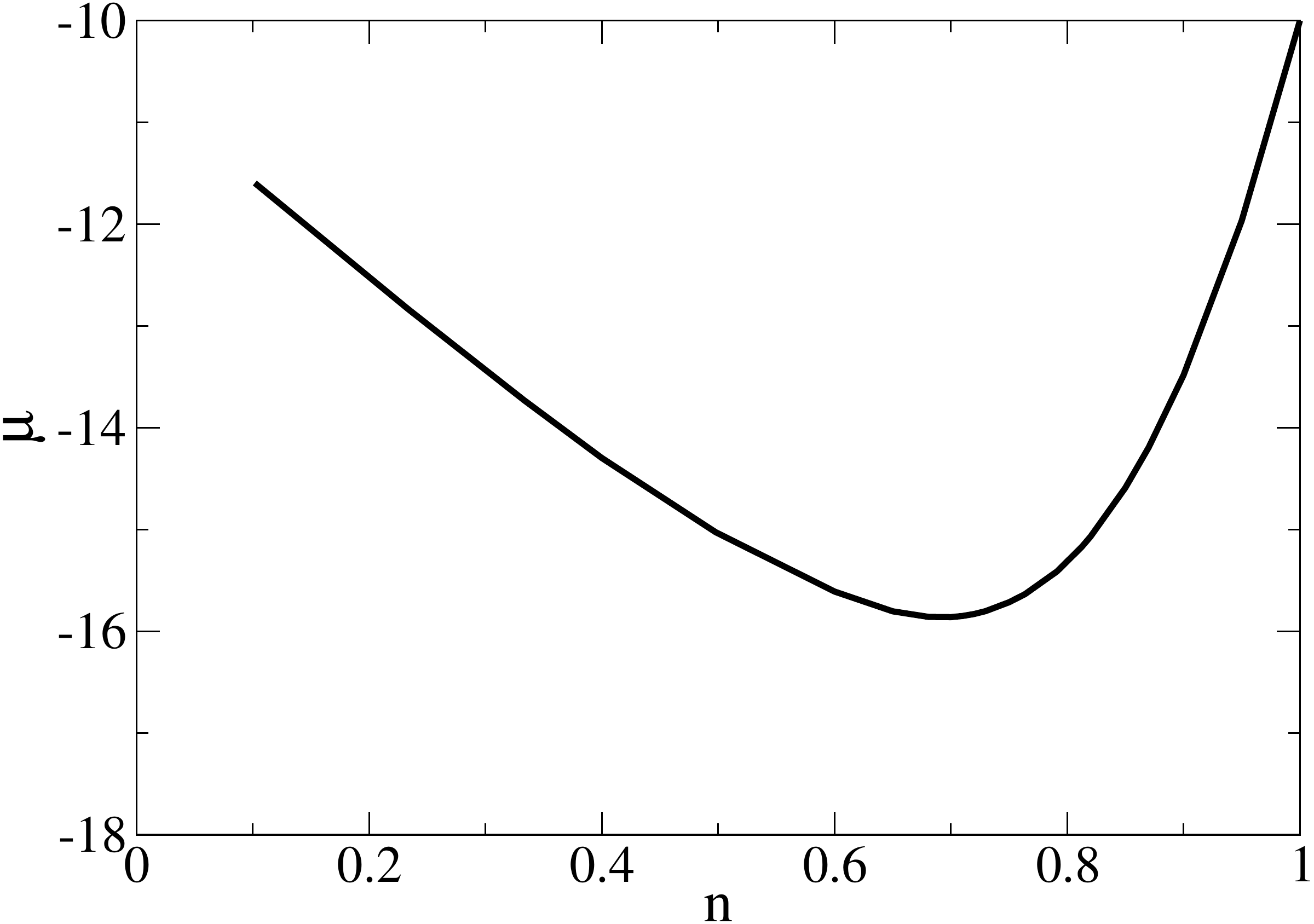}
\caption{The chemical potential $\mu$ plotted as a function of the filling $n$ for $U/t=20.0$. Note that the slope is negative upto $n\simeq 0.7$ indicating a negative compressibility. The range of fillings for which $dn/d\mu<0$ increases with $U/t$, so that eventually for very large couplings the system is unstable for all fillings.}
\label{muvsn}
\end{figure}
The problem with this diagrammatic approach is that it predicts an unphysical negative compressibility in the BEC limit. In Fig. (\ref{muvsn}) we have plotted $\mu$ as a function of $n$ for $U/t=20.0$. Clearly, the slope of $\mu$ versus $n$ is negative for a large range of $n$ indicating negative compressibility. However, we know that in this limit the system is a lattice Bose-gas with a hard-core repulsion coming from Pauli exclusion and a nearest neighbor repulsion proportional to $t^2/U$. Hence the system is stable in the BEC limit and the negative compressibility within the diagrammatic approach is therefore an unphysical result.

\section{Details of large-$N$ formalism}
\label{largeN_details}
The thermodynamic properties of the system can be obtained from the partition function in the grand-canonical ensemble $Z(\mu, \beta)$, where $\beta^{-1}$ is the temperature $T$ of the system. Indeed, $Z$ is related to the thermodynamic potential as $\Omega(\mu, \beta) = - \beta^{-1} \ln Z$.
This partition function can be expressed as a Feynman path integral over Grassmann fields ${\bar \Psi}_{\alpha\sigma}$ and $\Psi_{\alpha\sigma}$
\begin{equation}
Z=\int D{\bar \Psi}_{\alpha\sigma} D\Psi_{\alpha\sigma} \exp(-S_{\Psi})
\end{equation}
with the action in imaginary time $\tau$
\begin{equation}
S_{\Psi}=\int_0^{\beta}d\tau \sum_{i\alpha\sigma}\left( {\bar \Psi}_{i\alpha\sigma}(\tau)\partial_{\tau}\Psi_{i\alpha\sigma}(\tau)+H[{\bar \Psi}_{i\alpha\sigma}, \Psi_{i\alpha\sigma}]\right).
\end{equation}
The quartic fermionic interaction term in the Hamiltonian can be decoupled by introducing a
Hubbard-Stratonovich field $\Delta(x)$ at each $x = ({\bf x}_i, \tau)$ which couples to $\sum_{\alpha} \bar{\Psi}_{i\alpha\uparrow}(\tau)\bar{\Psi}_{i\alpha\downarrow}(\tau)$.
The partition function can then be written as 
$
Z=\int D\Delta D\Delta^{*}D{\bar \Psi}_{i\alpha\sigma} D\Psi_{i\alpha\sigma} \exp(-S_{\Psi,\Delta})$ 
with a full action
\begin{eqnarray}
S_{\Psi,\Delta} = &&\int d\tau \sum_i\Big ( { N \mid \Delta (x)\mid ^2 \over U}\\
& &- \int d\tau' \sum_{i j\alpha} \psi^\dagger_{i,\alpha} (\tau) {\bf G}^{-1}_{ij}(\tau, \tau') \psi_{j,\alpha} (\tau') \Big ),\nonumber
\end{eqnarray}
where we have introduced the Nambu spinors $\psi^\dagger_{i\alpha}(\tau) = (\bar{\Psi}_{i\alpha\uparrow}(\tau), \Psi_{i\alpha\downarrow}(\tau))$. The inverse Nambu-Gorkov Green's function $ {\bf G}^{-1}_{ij}(\tau, \tau')$ is given by
\begin{eqnarray}
\left(
\begin{array}{cc}%
        (-\partial_{\tau}+\mu)\delta_{i,j} + t \delta_{<i,j>}  & \Delta(x)\delta_{i,j} \\
        \Delta^*(x)\delta_{i,j} &( -\partial_{\tau}-\mu)\delta_{i,j} -t\delta_{<i,j>}
\end{array}
\right) \nonumber \\ \times \delta(\tau-\tau')\,\,\,\,
\label{nambuG}
\end{eqnarray}
with the notation $\delta_{<i,j>} = 1$ only if the $i$ and $j$ sites are nearest neighbors and zero otherwise. The functional integral is now both Gaussian in the fermionic fields and diagonal in the flavor index $\alpha$.  After integrating over these 
 Grassmann variables we get 
\begin{equation}\label{partition_function_Delta}
Z = \int D\Delta D\Delta^{*} \exp(-S_\Delta)
\end{equation}
with
an effective action $S_{\Delta}$ which only depends on the auxiliary fields $\Delta(x)$ in the form
\begin{equation}
S_\Delta = N\int dx\, {\mid \Delta(x)\mid ^2 \over U} - N \int dx \,  \Tr \ln \G^{-1}[\Delta(x)]
\label{S_Delta_x}
\end{equation}
where $\int dx = \sum_i \int d\tau$.

Assuming that the saddle-point auxiliary field is space- and time-independent (i.e. $\Delta(x) = \Delta_0$), the thermodynamical potential $\Omega$ is of the form $\Omega(\mu, \beta) \simeq N\Omega_0= S_\Delta(\Delta(x) = \Delta_0) / \beta$. Fluctuations around the saddle point yield corrections that are smaller than this term by powers of $1/N$; thus the full thermodynamic potential will be expanded in the form
\begin{equation}
{\Omega \over N} = \Omega_{0} + {1\over N} \Omega_g + \cdots 
\label{Omega_N_expansion}
\end{equation}

\subsection{Saddle point approximation - Mean field theory at T=0}

To find the uniform, static saddle point of the effective action $S_\Delta$, we replace $\Delta(x)$ by the space-time independent quantity
$\Delta_0$. 
Fourier transforming all the fields to the reciprocal (momentum) lattice and Matsubara frequencies, 
the effective action is given by
\begin{equation}
S_{\Delta}[\Delta_0]=N\frac{\beta \Delta_0^2}{U}-N\sum_{{\bf k}, ik_n} \operatorname{tr} \ln {\bf G}_0^{-1}(k) \equiv NS_0
\label{swrong}
\end{equation}
with 
\begin{equation}
{\bf G}_0^{-1}(k)= \left( \begin{array}{cc}
ik_n-\xi_{\bf k} & \Delta_0 \\
\Delta_0 & ik_n+\xi_{\bf k} \\
\end{array} \right)
\end{equation}
where $ik_n=(2n+1)\pi i/\beta$ are the fermionic Matsubara frequencies. The saddle point condition is~\cite{saddle point comment} $d S_0 / d \Delta_0 = 0$, which can be rewritten as
\begin{equation}
\frac{1}{U}=\sum_{\bf k}\frac{1}{2E_{\bf k}} \label{gap} 
\end{equation}
where $E_{\K}=\sqrt{\xi_{\K}^2+\Delta_0^2}$. 
%
The thermodynamic potential in the mean-field approximation is then 
\begin{equation}
\Omega_0 =S_0/\beta=\frac{\Delta_0^2}{U}-\sum_{\bf k}(E_{\bf k}-\xi_{\bf k})
\label{wrongomega}
\end{equation}

The mean field number equation can be obtained from 
\begin{equation}
\left(\frac{\partial \Omega_0}{\partial \mu}\right)_{T, V}=-n \,\,\,\,\,\,\textup{or}\,\,\,\,\,\, n=\sum_{\bf k}\left(1-\frac{\xi_{\bf k}}{E_{\bf k}}\right) \label{number}
\end{equation}
Eqs. (\ref{gap}, \ref{number}) must be solved self-consistently 
to obtain the mean field gap parameter $\overline{\Delta_0}$ corresponding to the mean field chemical potential $\overline{\mu}$, as well as finding the chemical potential which yields the desired density $n$.

\subsection{Gaussian fluctuations at $T=0$}
In order to go beyond the mean field approximation we must consider perturbations of the auxiliary field $\Delta(x)$ beyond the saddle-point, in the form
\begin{equation}
\Delta(x)=\Delta_0+\eta(x)
\end{equation}
where the complex bosonic field $\eta(x)$ describes space-time dependent fluctuations around the uniform static
value $\Delta_0$. We next expand the action $S_{\Delta}$ in Eq. (\ref{S_Delta_x}) to quadratic order in $\eta$, using that the saddle point condition (\ref{gap}) ensures that there is no term linear in $\eta$. Thus, the action is of the form $S_{\Delta}=NS_0+S_g+...$ with a Gaussian piece of the form
\begin{equation}\label{S_g_formula}
S_g=\frac{1}{2N}\sumq (\eta^*(q) \eta(-q))\,\M(q)\left( \begin{array}{c}
\eta(q) \\
\eta^*(-q)
\end{array} \right)
\end{equation}
where $iq_l=i2\pi l/\beta$ are the Bose-Matsubara frequencies and the matrix elements of the inverse fluctuation
propagator $\M$ are given by
\begin{eqnarray}
\M_{11}(q)=\M_{22}(-q)=\frac{1}{U}+\sumk \G^0_{22}(k)\G^0_{11}(k+q) \hspace{.2in} \\ \nonumber
=\frac{1}{U}+\sum_{\K}\left(\frac{u_{\K}^2u_{\Kp}^2}{iq_l-E_{\K}-E_{\Kp}}-\frac{v_{\K}^2v_{\Kp}^2}{iq_l+E_{\K}+E_{\Kp}} \right)
\end{eqnarray} and
\begin{eqnarray}
\M_{12}(q)=\M_{21}(q)=\sumk \G^0_{12}(k)\G^0_{12}(k+q) \hspace{.8in} \\ \nonumber
=\sum_{\K}u_{\K}u_{\Kp}v_{\K}v_{\Kp}\left(\frac{1}{iq_l+E_{\K}+E_{\Kp}}-\frac{1}{iq_l-E_{\K}-E_{\Kp}} \right)
\label{M}
\end{eqnarray}
Here we use the standard BCS notation $u_{\K}^2=1-v_{\K}^2=(1/2)(1+\xi_{\K}/E_{\K})$ and $\Kp=\K+\Q$. 

Writing the partition function upto Gaussian order
\begin{equation}\label{Z_gaussian}
Z\simeq \exp(-NS_0)\int D\eta D\eta^{\dagger}\exp(-S_g)
\end{equation}
and integrating out the Gaussian fluctuations we obtain (see Appendix \ref{converence_appendix} for details) the Gaussian contribution to the thermodynamic potential 
\begin{eqnarray}\label{omegag}
\Omega_g &=& \frac{1}{2\beta}\sum_{iq_n,{\bf q}}\ln\left( \frac{{\bf M}_{11}}{{\bf M}_{22}}\operatorname{Det}{\bf M}(q)\right)e^{iq_n0+}\\ \nonumber
&=&\frac{1}{2\beta}\sumq \ln \left(U^2 \Det \M(q)\right)+\frac{U}{2}\sum_{\K}(u_\K^2-v_\K^2).
\end{eqnarray}

In a previous article~\cite{rsr} some of us showed that this Gaussian fluctuation contribution 
can be physically interpreted by analytically continuing the bosonic Matsubara frequency to the real axis $iq_l \rightarrow z = \omega + i0^+$. We are thus led to the study of the analytical properties of $\ln {\rm Det}\, {\bf M} (\Q, z)$. The zeroes of $ {\rm Det}\, {\bf M} (\Q, z = \omega_0(\Q))$ (which correspond to poles of the fluctuation propagator ${\bf M}^{-1}$) correspond to the frequencies $\omega_0(\Q)$ of collective excitations of the system with momentum $\Q$. These excitations are the $\Q \rightarrow 0$ Goldstone modes of the order parameter in the broken symmetry superfluid state. Additionally, the fluctuation propagator has 
branch cuts on the real axis originating at $E_c(\Q)=\pm\operatorname{min}(E_{\K}+E_{\K+\Q})$. These branch cuts represent the two-particle continuum of states for scattering of gapped quasiparticles. The Gaussian contribution (\ref{omegag}) can be then rewritten as 
\begin{equation}
\Omega_g = {1\over 2} \sum_\Q \left [ \omega_0 (\Q) - E_c(\Q) - \int_{-\infty}^{-E_c({\bf q})}{d\omega \over \pi} \delta({\Q}, \omega) \right ] + {\cal R}
\end{equation}
where the last integral describes the contribution of the virtual scattering of quasiparticles with a phase shift $\delta(\Q, \omega)$ whose particle continuum begins at $E_c(\Q)$ and the last term $\cal R$ comes from using the correct convergence factors in the calculation (see Appendix B).

\begin{figure}[tbp]
\includegraphics[width=3in]{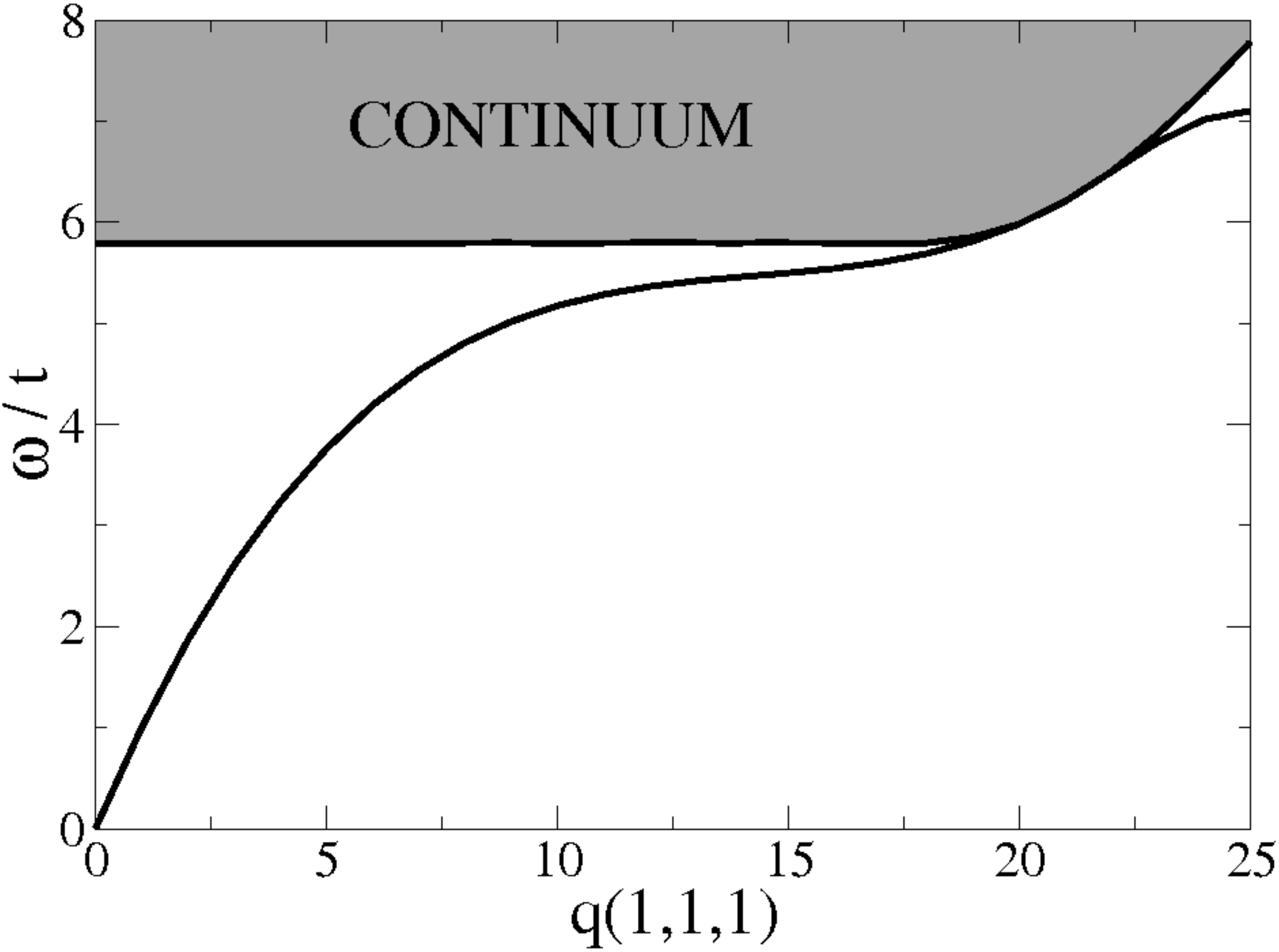}
\caption{Spectrum of excitations which contribute to the leading $1/N$ corrections to the thermodynamic potential plotted at unitarity for $n=0.5$ and along the main diagonal ($q_x=q_y=q_z$) of the Brillouin zone of a $50\times50\times50$ lattice. The solid line is the collective sound mode given by $\Det\M=0$. The shaded region is the two-particle continuum given by the branch cut of the fluctuation propagator.}
\label{collective}
\end{figure}

To illustrate this excitation spectrum we plot in Fig. \ref{collective} the two particle continuum and the collective excitations along the main diagonal $q\,(1,1,1)$ of
the Brillouin zone, at unitarity and for $n=0.5$. 
For small $\Q$, the collective excitation spectrum is linear
indicative of sound modes, eventually hitting the two-particle continuum. 

In the BCS limit, the contribution of the collective mode is negligible due to phase space restrictions and the two-particle continuum dominates. In the BEC limit, the two-particle continuum lies at a much higher energy scale and the
low-energy excitations are entirely given by the gapless sound modes. Further, at half-filling, one would expect the collective excitation spectrum to be gapless at $\Q=(\pi,\pi,\pi)$ indicating new Goldstone modes due to the onset of CDW order \cite{arun}. However,
since we only decouple the quartic interaction in the p-p channel, we do not see the CDW order and hence there is no softening of $(\pi,\pi,\pi)$ mode at half-filling within our theory.

\subsection{Corrections of order $1/N$}
In order to calculate the leading order corrections to the thermodynamical quantities, such as the chemical potential in this case, we write it as the expansion
\begin{equation}
\mu =\overline{\mu}+\frac{\delta\mu}{N}+\dots .
\end{equation} 
Naturally, given that the gap parameter $\Delta_0$ is a function of $\mu$, it will also have an expansion in powers of $1/N$ derived from this expansion. Indeed, 
\begin{eqnarray}
\Delta_0 &=& \overline{\Delta_0} + {1\over N} \delta\Delta_0;\\
\delta\Delta_0 &=& {d\Delta_0 \over d\mu}\, \delta\mu = - {(\partial^2 S_0 / \partial \mu \partial \Delta_0) \over
(\partial^2 S_0 / \partial \Delta_0^2) } \, \delta\mu.
\end{eqnarray}

Next, expanding the number equation to linear order in $1/N$ and remembering that in calculating derivatives with respect to $\mu$ (which we denote here as  $d/d\mu$) the parameter $\Delta_0$ actually changes with $\mu$, we get
\begin{eqnarray}
-n=\left(\frac{d\Omega}{d\mu}\right)=\left(\frac{d\Omega_0}{d\mu} \right )\\ \nonumber +\frac{1}{N}\left[\left(\frac{d^2\Omega_0}{d\mu^2}\right)
\delta\mu+\left(\frac{d\Omega_g}{d\mu}\right)\right]
\end{eqnarray}
which yields
\begin{equation}
\delta\mu=-\left(\frac{d\Omega_g}{d\mu}\right)/\left(\frac{d^2\Omega_0}{d\mu^2}\right)
\label{deltamu}
\end{equation}
where all quantities are evaluated at the mean field value $\mu = \overline{\mu}$.


\section{Convergence scheme in the $1/N$ expansion}
\label{converence_appendix}
In this appendix we develop a convergence scheme for including the effects of $1/N$ corrections to $\Omega$. The Gaussian part of the thermodynamic potential is given by
\begin{equation}
\Omega_g=\frac{1}{2\beta}\sumq \ln(\M_{11}(q)\M_{22}(q)-\M_{12}^2(q))
\label{1}
\end{equation}
where 
\begin{eqnarray}
\M_{11}(q)=\frac{1}{U}+\sum_{\bf k} \left( \frac{u^2 u'^2}{iq_l-E-E'}-\frac{v^2 v'^2}{iq_l+E+E'} \right) \nonumber \\
\M_{12}(q)=\sum_{\bf k}uvu'v'\left( \frac{1}{iq_l+E+E'}-\frac{1}{iq_l-E-E'} \right) \nonumber
\end{eqnarray}
The expression for $\Omega_g$ in Eq. (\ref{1}) is ill-defined in the absence of convergence factors. We remind ourselves here that $\M_{11}$ and $\M_{22}$ have different convergence
factors. We next split $\ln[\M_{11}\M_{22}-\M_{12}^2]$ as
\begin{equation}
\ln(\M_{11})e^{+iq_l0^+}+\ln(\M_{22})e^{-iq_l0^+}+\ln\left(1-\frac{\M_{12}^2}{\M_{11}\M_{22}}\right)
\end{equation} 
Note, the last term does not need a convergence factor. We next note that $\M_{22}(-q)=\M_{11}^{\ast}(q)$ and so
upon summation over positive and negative values of $q$,
$\M_{22}$ can be combined with $\M_{11}$ to give
\begin{equation}
\Omega_g=\frac{1}{2\beta}\sumq \ln\left(\frac{\M_{11}}{\M_{22}}\Det\M(q)\right)e^{+iq_l0^+}
\label{2}
\end{equation}
We have now endowed $\Omega_g$ with a single convergence factor and in order to remove the log divergence at
large $U$ we now simply add to $\Omega_g$ a term $\sum_{iq_l}\ln(U^2) e^{+iq_l0^+}$, which has no poles or singularities in
the left-half plane and therefore contributes nothing to the Matsubara sum except rendering $\Omega_g$ finite in
the limit $U\rightarrow\infty$.

We thus redefine $\M\rightarrow U\M$ and split the sum in Eq. (\ref{2}) into two parts:
\begin{equation}
\Omega_g=\frac{1}{2\beta}\sumq \left[\ln\left(\frac{\M_{11}}{\M_{22}}\right)e^{+iq_l0^+}+\ln\Det\M\right]
\label{3}
\end{equation}
Again, the last term in Eq. (\ref{3}) is manifestly convergent and hence does not require a convergence
factor. However, the first term is ultra-violet divergent since in the large $iq_l$ limit it goes like
\begin{equation}
\frac{\M_{11}}{\M_{22}} \sim 1+\frac{2\alpha(\Q)}{iq_l},
\end{equation}
where $\alpha(\Q)=U\sum_{\K}(u^2u'^2-v^2v'^2)$. To regulate the offending term, we subtract and add a term $\sumq
\alpha(q)[1/(iq_l+a) + 1/(iq_l-a)]e^{+iq_l0^+}$, where $a$ is any real number to obtain
\begin{eqnarray}
\hspace{-.2in} \frac{1}{2\beta}\sum_{\Q,
iq_l}\left[\ln(\frac{\M_{11}}{\M_{22}})-\alpha(q)\left(\frac{1}{iq_l+a}+\frac{1}{iq_l-a}\right)\right]e^{+iq_l0^+}\\\nonumber + \frac{1}{2\beta}\sumq
\alpha(q)\left(\frac{1}{iq_l+a}+\frac{1}{iq_l-a}\right)e^{+iq_l0^+}
\end{eqnarray}
Now the first term is explicitly convergent and hence the convergence factor can be dropped. However, without the
convergence factor, the first term is an odd function of $q_l$ and hence the Matsubara sum gives zero. So, we are left with
only the second term, where the Matsubara sum can again be converted into an integral along the imaginary axis. We
can further analytically continue on the left half of the complex plane (contribution from right half is zero at
$T=0$ due the Bose distribution function) and close the contour counter-clockwise to enclose the only singularity
at $z=-a$. The contour integration gives $(U/2)\sum_{\K}(u^2-v^2)$ for the second term and thus the correct form of $\Omega_g$ is given
by
\begin{equation}
\Omega_g=\frac{1}{2\beta}\sumq \ln \Det \M(q)+\frac{U}{2}\sum_{\K}(u^2-v^2)
\end{equation}
In the next appendix (Appendix \ref{num_eval_appendix}) we outline the numerical steps for evaluating $\Omega_g$.

\section{Numerical evaluation of $\Omega_g$}
\label{num_eval_appendix}
The first step in the calculation of the Gaussian part of the thermodynamic potential is to solve the gap
equation for a given chemical potential. Since we do not know the analytical form of the
number equation once we include Gaussian fluctuations, we work in the grand canonical ensemble and obtain
$\Delta(\mu)$ from equations (\ref{gap}). We next numerically compute $\Omega_g[\mu,
\Delta(\mu)]$ using the formula in equation (\ref{omegag}). All the $3$ momenta sums are over the
entire Brillouin zone for a $20\times 20\times 20$ lattice and have an implicit factor of total number of lattice sites in front.
The Matsubara sum over the imaginary frequencies $iq_n$ is computed along the imaginary axis for each $\Q$ mode.
The integral in equation(\ref{omegag}) is split as follows:
\begin{equation}
\int_0^{\infty}\ln(\Det \M)dy=\int_0^{y_c}\ln(\Det \M)dy+\int_{y_c}^{\infty}F(y)dy
\end{equation}
where the first integral on the left hand side is computed numerically and the second integral is evaluated
analytically using the large $y$ asymptote of the integrand. The function $F(y)$ is given by
\begin{equation}
F(y)=\frac{4U}{y^2}\sumk (u^2u'^2+v^2v'^2)(E+E')
\end{equation}
Here one has to be careful about the integrable log-divergence at $\Q=(0,0,0), y=0$ coming from Goldstone's Theorem. To take this into account we expand the integrand for $\Q=(0,0,0)$ and small $y$ and obtain $\ln(\Det\M({\bf 0},y)) \approx \ln(K y^2)$, where $K=a^2+b^2-g^2$ and
\begin{eqnarray}
a^2 &=& 2U\left(1-U\sum_{\K}\frac{u^4+v^4}{2E}\right)\left(\sum_{\K}\frac{u^4+v^4}{8E^3}\right) \\ \nonumber
b^2&=&\left[U\sum_{\K}\frac{-u^4+v^4}{4E^2}\right]^2 \\ \nonumber
g^2&=&-2U\sum_{\K}\frac{u^2v^2}{4E^3}
\end{eqnarray}
We note that the terms independent of $y$ in the expressions for $a, b$ and $g$ cancel due to Goldstone's theorem
and the term linear in $y$ cancel due to symmetry. The integrand $\ln(\Det\M({\bf 0},y))$ for $\Q={\bf 0}$ is then integrated
between limits $0$ and a small value of $y=y_s$. The rest of the integral for $\Q={\bf 0}$ is evaluated numerically
between $y_s$ and $y_c$, and analytically between limits $y_c$ and $\infty$ using the asymptotic form $F(y)$.

\section{Comparison of our $1/N$ expansion at $T=0$ with \cite{vsr} (VSR)}
\label{vsr_appendix}
In this appendix we show the equivalence of our method for obtaining the $1/N$ corection to the chemical potential at $T=0$ with the one obtained by \cite{vsr} (VSR). We start with equation (\ref{deltamu}) and write the numerator as 
\begin{equation}
\frac{d\Omega_g}{d\mu}=\frac{\partial\Omega_g}{\partial\mu}+\frac{\partial\Omega_g}{\partial\Delta_0}\frac{d\Delta_0}{d\mu}
\label{num}
\end{equation}
We next evaluate the denominator of equation (\ref{deltamu})
\begin{equation}
\frac{d^2\Omega_0}{d\mu^2}=\frac{\partial^2\Omega_0}{\partial\mu^2}+\left(\frac{\partial^2\Omega_0}{\partial\Delta_0\partial\mu}\right)\frac{d\Delta_0}{d\mu}
\label{den}
\end{equation}
Putting equations (\ref{num}, \ref{den}) in equation (\ref{deltamu}) we obtain
\begin{eqnarray}
\delta\mu=-\left[\left(\frac{\partial\Omega_g}{\partial\mu}\right)\left(\frac{\partial^2\Omega_0}{\partial\Delta_0^2}\right)+\left(\frac{\partial\Omega_g}{\partial\Delta_0}\right)\left(\frac{\partial^2\Omega_0}{\partial\Delta_0\partial\mu}\right)\right]/ \nonumber \\ \left[\left(\frac{\partial^2\Omega_0}{\partial\mu^2}\right)\left(\frac{\partial^2\Omega_0}{\partial\Delta_0^2}\right)-\left(\frac{\partial^2\Omega_0}{\partial\Delta_0\partial\mu}\right)\right]\,\,\,\,
\label{same}
\end{eqnarray}
which is the same as equation (3.31) of \cite{vsr}. However, it should be emphasized that the respective $1/N$ corrections to the gap parameter within our theory and within VSR are nevertheless different and can be traced back to the question of feedback discussed earlier. 

\section{Details of the superfluid density calculation for large-$N$}
\label{rhos_details}
Here we extend our large-$N$ formalism to the calculation of the superfluid density~\cite{Taylor} in a
lattice. We start by putting a (time-independent) phase twist $\theta(x) = {\bf Q} \cdot {\bf x}_i = \theta_i$ on the order parameter
\begin{equation}
\Delta(x)\rightarrow e^{i\theta_i}\Delta(x)
\end{equation}
in the expression (\ref{nambuG}) for the Nambu-Gorkov propagator, which is used to calculate the full action $S_{\Psi, \Delta}$. We shall assume that the phase difference is a constant $\theta$ for any two neighboring sites along the (arbitrarily chosen) $x$-direction; thus, we have the relation $Q = Q_x = \theta /a$ where $a$ is the lattice spacing and the superfluid density is
\begin{equation}
n_s=4ma^2\left(\frac{\partial^2{\Omega}}{d\theta^2}\right)_\mu,
\label{rhos2}
\end{equation}
where the mass in the lattice is the combination $m = 1/(2ta^2)$, i.e. the effective mass for fermions at the bottom of the band.

We can remove the phase twist from $\Delta(x)$ by applying 
a unitary transformation to the Grassmann fields of the form
\begin{equation}
\psi_{i\alpha} (\tau) = {\bf U}_i \, \psi_{i\alpha} (\tau)
\end{equation}
with 
\begin{equation}
{\bf U}_i = \left (
\begin{array}{cc}
e^{-i\theta_i/2} & 0\\
0& e^{i\theta_i/2} 
\end{array}
\right ).
\label{U matrix}
\end{equation}
so that the inverse Green's function in (\ref{nambuG}) now becomes ${\bf G}^{-1}_{ij}(\tau, \tau') \rightarrow {\bf U}_i {\bf G}^{-1}_{ij}(\tau, \tau') {\bf U}_j^\dagger$. As can be easily verified, this leaves the form of ${\bf G}^{-1}_{ij}(\tau, \tau')$ unchanged from (\ref{nambuG}) except for the hopping term, which gains the phase difference $t \rightarrow t \, \exp(\pm i (\theta_j - \theta_i)/2)$ in the first (second) diagonal element. Transforming to the reciprocal lattice, we obtain an 
%
%
inverse Green's function of the form
\begin{equation}
{\tilde \G}_0^{-1}=\left( \begin{array}{cc}
(ik_n-\xi_{{\bf k} + {\bf Q}/2})e^{-ik_l0^+} & \Delta_0 \\
\Delta_0 & (ik_n+\xi_{{\bf k} - {\bf Q}/2})e^{+ik_l0^+}
\end{array} \right)
\end{equation}
Taking the limit of small $\theta$, we see that this corresponds to shifting the Matsubara frequencies and the energy dispersion respectively as 
\begin{eqnarray}
i{\tilde k}_n&=&ik_n- {\theta \over 2a} (\partial
\epsilon_{\K}/ \partial k_x) \\ 
\nonumber {\tilde \xi_{\K}}&=& 
\xi_{\K} + \frac{\theta^2}{8a^2}\frac{\partial^2 \epsilon_{\K}}{\partial k_x^2}
\label{split}
\end{eqnarray}

The effective action at a fixed $\theta$, from which the saddle point condition is derived, satisfies
\begin{eqnarray}\label{omega0tilde}
{S_0(\Delta_0; \mu, \theta)\over N \beta}  &=& \frac{\Delta_0^2}{U}-\frac{1}{\beta}\sumk\Tr \ln {\tilde \G}_0^{-1}(k)\\
&=& {S_0(\Delta_0; \mu, \theta=0)\over N \beta} + {\theta^2 \over 8 a^2} \sum_{\bf k} (1-{\xi_{\bf k} \over E_{\bf k}}){\partial^2 \epsilon \over \partial k_x^2}\nonumber
\end{eqnarray}
where we have expanded the energy dispersions to quadratic order in $\theta$ and used the fact that the shift in the Matsubara frequencies is not important once they are summed over as long as $\theta$ is small. The saddle point condition $\delta S_0 / \delta \Delta_0 = 0$ yields the small $\theta$ expansion $\Delta_0 (\mu, \theta) = \Delta_0(\mu, 0) + \alpha(\mu) \theta^2$, where 
\begin{equation}
\alpha(\mu)= -{3\over 8\Delta_0 a^2} \frac{\sum_{\bf k} (\partial^2 \epsilon_{\bf k} / \partial k_x^2)\xi_{\bf k}/E_{\bf k}^3}{\sum_{\bf k} (1/E_{\bf k}^3)}
\end{equation} 
and all quantities are evaluated at $\theta = 0$. 

\subsection{Mean Field Superfluid Density}

The mean field thermodynamic potential per flavor at $T=0$ for a system with a phase twist is given by
\begin{equation}
{{\Omega}_0(\mu, \theta)}= {1\over N \beta}S_0(\Delta_0(\mu, \theta); \mu, \theta) 
\end{equation}
Using that the $\Delta_0$ dependence on $\theta$ is obtained from the saddle point condition $\delta S_0 /\delta \Delta_0 = 0$ we can obtain the superfluid density as
\begin{eqnarray}
n_s^0 &=& {1\over N\beta} \left( {\partial^2 S_0 \over \partial \theta^2} \right )_{\mu, \Delta}\\
&=& \sum_{\K}\left( 1-\frac{\xi_{\K}}{E_{\K}} \right ) \cos(k_xa)
\end{eqnarray}

\subsection{Calculation of $n_s$ including Gaussian fluctuations}

In order to include Gaussian fluctuations, we need to calculate the Gaussian part of the action $S_g(\Delta_0; \theta, \mu)$ in the presence of the phase twist $\theta$. The inclusion of the effects of Gaussian fluctuations in the calculation of the superfluid density follows the same methodology used in section \ref{Results section}. The thermodynamic potential per flavor to first order in $1/N$ is of the form
\begin{equation}
\frac{\Omega(\mu, \theta)}{N} = \Omega_0(\mu, \theta) + {1\over N\beta} \Omega_g(\mu, \theta).
\end{equation}
where again $\Omega_g(\mu, \theta) = S_g(\Delta_0(\mu, \theta); \mu, \theta)/\beta$.

Before we give an explicit expansion of the superfluid density in orders of $1/N$ on a lattice, let us go back to the continuum limit and outline the calculation of the $1/N$ corrections to $n_s$ in the continuum. This would hopefully elucidate some of the technical points differentiating a lattice calculation from the continuum. To this effect we prove that for a translationally invariant system, the relation $n_s=n$, is respected even at the $1/N$ level. For an energy dispersion $\epsilon_{\K}=\K^2/2m$, the shift in single
particle energies due to the introduced twist (see second line in Eq. \ref{split}) is only a constant and can be incorporated into
the chemical potential. Since, the phase twist is uniformly distributed across the system the order parameter transforms as: $\Delta(x) \rightarrow \Delta(x)e^{i {\bf Q}.{\bf r}}$ and the Green's function transforms as
\begin{equation}
\G_{ij}(ik_n, \K, \mu)\rightarrow \G_{ij}(i{\tilde k}_n, \K, \mu-{\bf Q}^2/8m)
\end{equation}
where the $i{\tilde k}_n=ik_n-\K.{\bf Q}/2m$ are the Doppler shifted Matsubara frequencies. Then,
$\M_{11}(iq_l, \Q; {\bf Q})=1/U+\sumk \G_{22}(ik_n-\frac{\K.{\bf Q}}{2m}, \K; \mu-\frac{{\bf
Q}^2}{8m})\G_{11}(ik_n-\frac{{(\K+\Q).\bf Q}}{2m}+iq_l, \K+\Q; \mu-\frac{{\bf Q}^2}{8m})=\M_{11}(iq_l-\frac{\Q.{\bf Q}}{2m}, \Q; \mu-\frac{{\bf Q}^2}{8m})$.
The effects of the phase twist at $T=0$ is therefore to shift the contour of integration for the Matsubara sum by an amount
proportional to $Q$ along the real axis and to shift the chemical potential $\mu$ by a constant amount ${\bf Q}^2/8m$.
In the limit $Q \rightarrow 0$, the shift of the contour of integration to the right keeps the Matsubara sum invariant and hence the phase twist
enters the thermodynamic potential only through a shift of the chemical potential. This means that the saddle point
condition in presence of the phase twist remains unchanged from the one in absence of
the same. Further, ${\Omega(Q)}$ only contains terms in powers of $Q^2$ and therefore we obtain,
\begin{equation}
4m\left(\frac{\partial^2{\Omega_g}}{\partial Q^2} \right)_{\overline{\mu}, Q\rightarrow 0}=-\left(\frac{\partial
\Omega_g}{\partial \mu}\right)_{\overline{\mu}}
\end{equation}
By the same logic, $4m(\partial {\Omega_0}/\partial Q^2)_{\overline{\mu}, Q\rightarrow 0}=-(\partial \Omega_0/\partial \mu)_{\overline{\mu}}$. Since, the number equation is given by $\partial (\Omega_0+\Omega_g/N)/\partial \mu = -n$ we obtain 
\begin{eqnarray}
n_s=4m\left(\frac{d^2{\Omega}}{dQ^2}\right)_{Q\rightarrow 0}&=&-\left(\frac{\partial\Omega_0}{\partial
\mu}\right)_{\overline{\mu}}-\frac{1}{N}\left(\frac{\partial\Omega_g}{\partial \mu}\right)_{\overline{\mu}} \nonumber \\
&=&n
\end{eqnarray}

We next consider the lattice and write the $1/N$ corrections to the thermodynamic potential as follows
\begin{eqnarray}
\Omega(\mu,\Delta_0(\mu,\theta),\theta)\,=&&\Omega_0(\mu,\Delta_0(\mu,\theta),\theta)\nonumber \\ &+&\frac{1}{N}\Omega_g(\mu,\Delta_0(\mu,\theta),\theta)
\label{2}
\end{eqnarray}
Further, $\mu=\overline{\mu}+\delta\mu/N$, which then combined with Eq. (\ref{rhos2}) yields the following expansion for the superfluid density
\begin{widetext}
\begin{eqnarray}
n_s&=&\left(\frac{\partial^2{\Omega}}{\partial \theta^2}\right)_{\mu}=\left(\frac{\partial^2{\Omega_0}}{\partial\theta^2}\right) + \frac{1}{N}{d \over d\mu}
\left(\frac{\partial^2{\Omega_0}}{\partial \theta^2}\right)\delta\mu
+\frac{1}{N}\left(\frac{\partial^2{\Omega_g}}{\partial\theta^2}\right)\nonumber\\ 
&=&n_s^0 +\frac{1}{N}\left[ 
{d \over d\mu}\left(\frac{\partial^2{\Omega_0}}{\partial \theta^2}\right)\delta\mu
+\left(\frac{\partial^2{S_g/\beta}}{\partial\theta^2}\right)_{\mu, \Delta_0}+\left(\frac{\partial{S_g/\beta}}{\partial\Delta_0}\right)_{\mu,\theta}
\left( 2 \alpha(\mu) \right)\right] 
\label{omegagtheta}
\end{eqnarray}
\end{widetext}
where $d/d\mu=\partial/\partial\mu+(\partial\mu/\partial\Delta_0)\partial/\partial\Delta_0$. Note the presence of an explicit $\partial/\partial\Delta_0$ derivative which was absent in the expression for $n_s^0$ because of the saddle point condition $\partial\Omega_0/\partial\Delta_0=0$. Eq. \ref{omegagtheta} gives the expression for $n_s$ that was used in section \ref{SFdensity}.

\section{Critical temperature using Hartree shifted NSR}
\label{hnsr_appendix}
In this appendix we shall use the Nozieres and Schmitt-Rink approach \citep{nsr} for calculating the critical temperature, with the modification that the single particle Green's function $\G_0$ now includes a Hartree shift which we call $\Sigma$. We shall work in the grand canonical ensemble at a fixed $\mu$ and hence $\Sigma$ is a function of $\mu$ and $T$. We approach the transition from above $T_c$ and look for the divergence of the t-matrix. This gives us a relation between $T_c$ and $\mu$   
\begin{equation}
\frac{1}{U}=\sum_{\K} \frac{1}{2\xi_{\K}}\tanh(\frac{\beta \xi_{\K}}{2})
\label{tc_gap_nsr}
\end{equation}
Here the Hartree shift is contained in $\xi_{\K}=\epsilon_{\K}-\mu+\Sigma$. Since we are working in the grand canonical ensemble the filling fraction would depend on the value of $\mu$ we choose. At a fixed temperature this dependence is given through the number equation
\begin{equation}
n=-\left.\frac{\partial\Omega(T,\mu)}{\partial\mu}\right|_{T}
\label{numeq_nsr}
\end{equation}
The Hartree shift, which depends on the filling fraction, is then given by 
\begin{equation}
\Sigma(\mu, T)=-n(\mu,T)U/2
\label{sigtc}
\end{equation}
In order to implement the number equation (\ref{numeq_nsr}) we proceed as follows:
For a given $U$ and $\mu$, we calculate $T_c$ and $\Sigma(\mu,T_c)$ by simultaneously solving Eqs. (\ref{tc_gap_nsr}) and (\ref{sigtc}). With these values of $T_c$ and $\Sigma$, we evaluate $\Omega(\mu, T=T_c)$. Next, keeping the temperature fixed we change $\mu$ to $\mu+\delta\mu$ and evaluate $\Sigma(\mu+\delta\mu,T=T_c)$ from Eq. (\ref{sigtc}). This lets us evaluate $\Omega(\mu+\delta\mu, T=T_c)$. The number equation can then be written in the form
\begin{equation}
n=-\frac{\Omega(\mu+\delta\mu, T=T_c)-\Omega(\mu, T=T_c)}{\delta\mu}
\end{equation}

\begin{figure}[tbp]
\begin{center}
\includegraphics[width=3in]{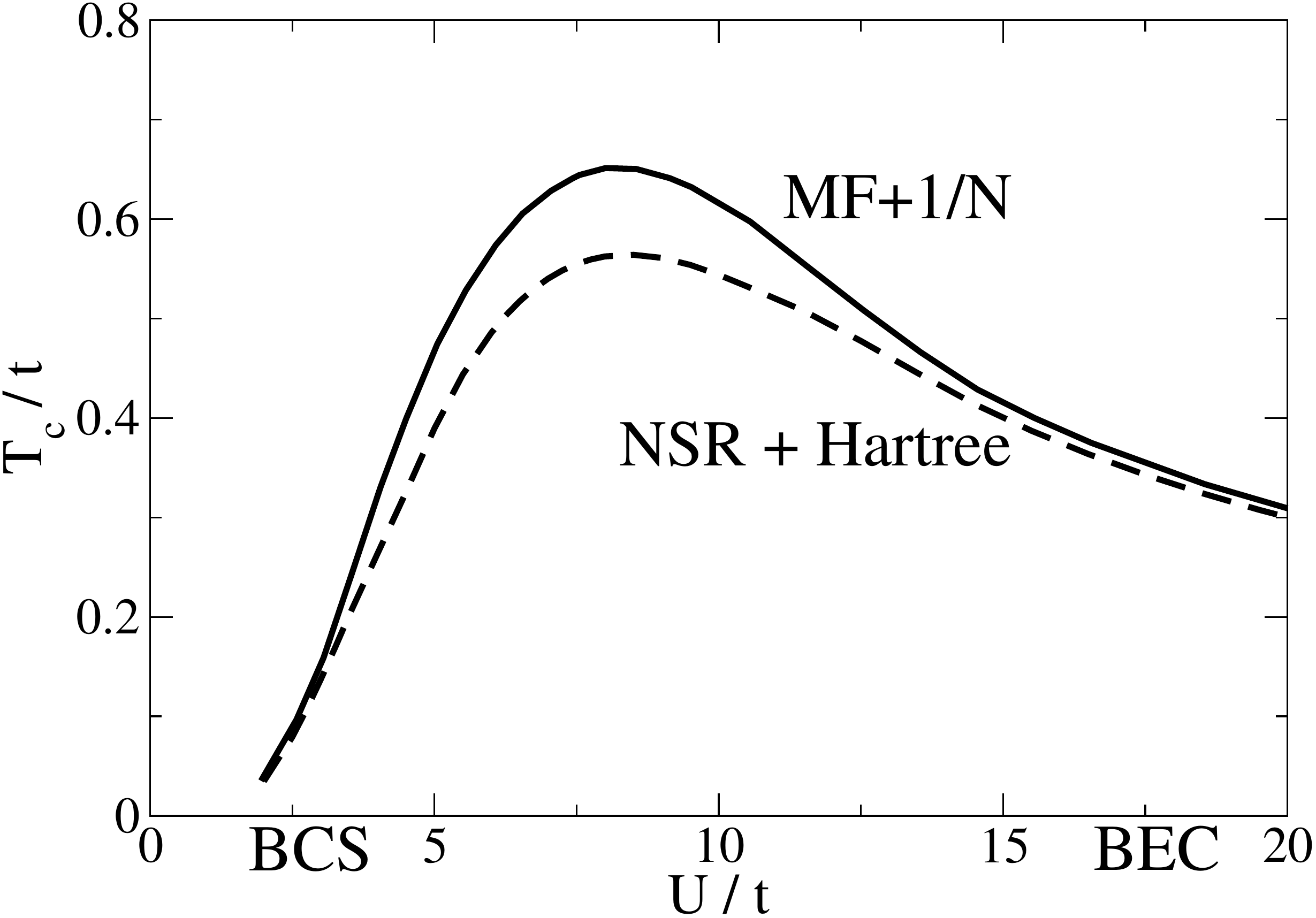}
\caption{Figure shows a comparison of the critical temperatures obtained within Hartree shifted NSR and large $N$ respectively. The filling fraction $n=0.5$. The maximum $T_c$ for the Hartree shifted NSR and the large $N$ theory can be compared to the results of Tamaki ${\it et\, al.}$ ($T_c^{\textup{max}}\simeq 0.66t$ for $n=0.5$) \cite{ohashi}}
\label{tchnsr}
\end{center}
\end{figure}

We next give an explicit formula for the thermodynamic potential $\Omega=\Omega_0+\Omega_g$. In presence of the Hartree shift, the mean field thermodynamic potential $\Omega_0$ is given by 
\begin{equation}
\Omega_0=-\frac{2}{\beta}\sum_{\K} \ln(1+e^{-\beta \xi_{\K}})+\frac{\Sigma^2}{U}
\end{equation}
As a check, note that setting $\Omega=\Omega_0$ in Eq. (\ref{numeq_nsr}) the above form for $\Omega_0$ gives us the familiar mean field number equation
\begin{equation}
n=\sum_{\K}\frac{2}{\exp(\beta \xi_{\K})+1}
\label{tc_n_nsr}
\end{equation}
To obtain the thermodynamic potential upto Gaussian order for
$T \geq T_c$, we note that $u_{\K}=1$ and $v_{\K}=0$ and hence $\M_{12}=\M_{21}=0$. Therefore
\begin{equation}
\Omega_g =\frac{1}{\beta}\sum_{\Q, iq_l} \left[ \ln(\M_{11}(q))-(\M_{11}(q)-1) \right]
\label{tc_omegag}
\end{equation}
where, we are justified to drop the convergence factor in the last line of Eq. (\ref{tc_omegag}) since
$(\M_{11}-1)$ takes out the leading $1/iq_l$ piece from
$\ln(\M_{11})$ (see Appendix \ref{diagram_appendix} for more details). For $T\geq T_c$, $\M_{11}(q)=1+U\sum_{\K}(1-f-f')/(iq_l-\xi-\xi')$, where $iq_l=i2\pi l/\beta$ are the Bose Matsubara frequencies and
$f(\xi)=1/[\exp(\beta \xi)+1]$ is the Fermi distribution function. We numerically calculate $\Omega_g$ using Eq. (\ref{tc_omegag}) and set up the number equation using the procedure outlined above. In order to obtain the $n$ dependence of $T_c$, the procedure is repeated for various values of $\mu$. The results are plotted in Figure (\ref{tchnsr}). 
We notice that there are quantitative differences between the HNSR and large-$N$ results. At this stage, we do not understand why the $T_c$ from HNSR is lower than the $T_c$ from large-$N$ theory.


\begin{thebibliography}{99}

\bibitem{leggett80}
A. J. Leggett, in {\em Modern Trends in the Theory of Condensed Matter},
edited by A. Pekalski and R. Przystawa (Springer-Verlag, Berlin, 1980).

\bibitem{eagles}
D. M. Eagles, Phys. rev. {\bf 186}, 456 (1969).

\bibitem{nsr}
P. Nozi\'{e}res and S. Schmitt-Rink, J. Low Temp. Phys. {\bf 59}, 195 (1985).

\bibitem{BECreview}
M. Randeria, in {\it Bose-Einstein Condensation},
edited by A. Griffin, D. Snoke, and S. Stringari, (Cambridge University Press, Cambridge, England, 1995), p.~355 - 392.

\bibitem{sademelo}
C. A. R. S\'{a} de Melo, M. Randeria, and J. R. Engelbrecht, Phys. Rev. Lett. {\bf 71}, 3202 (1993).

\bibitem{rsr} R.~Diener, R.~Sensarma, and M.~Randeria, Phys. Rev. A \textbf{77}, 023626 (2008).

\bibitem{engelbrecht97}
J. R. Engelbrecht, M. Randeria, and C. A. R. S\'a de Melo, Phys. Rev. B {\bf 55}, 15153 (1997).

\bibitem{Haussman}
R. Haussman, W. Rantner, S. Cerrito, and W. Zwerger, Phys. Rev. A {\bf 75}, 023610 (2007).

\bibitem{Sachdev}
P. Nikolic and S. Sachdev, Phys. Rev. A {\bf 75}, 033608 (2007).

\bibitem{vsr}
M. Y. Veillette, D. E. Sheehy, L. Radzihovsky, Phys. Rev. A {\bf 75}, 043614 (2007).

\bibitem{Carlson}
J. Carlson, S.-Y. Chang, V. R. Pandharipande, and K. E. Schmidt,
Phys. Rev. Lett. {\bf 91}, 050401 (2003);
S.-Y. Chang, V. R. Pandharipande, J. Carlson, and K. E. Schmidt,
Phys. Rev. A {\bf 70}, 043602 (2004).

\bibitem{Pieri-Strinati}
P Pieri, L. Pisani and G. C. Strinati,
Phys. Rev. B {\bf 72}, 012506 (2005).

\bibitem{jochim}
S. Jochim, M. Bartenstein, A. Altmeyer, G. Hendl, S. Riedl, C. Chin, J. Hecker Denschlag, and R. Grimm, Science {\bf 302}, 2101 (2003).

\bibitem{jin}
C. A. Regal, M. Greiner, and D. S. Jin, Phys. Rev. Lett. {\bf 92}, 040403 (2004).

\bibitem{ketterle}
M. W. Zwierlein, C. A. Stan, C. H. Schunck, S. M. F. Raupach, A. J. Kerman, and
W. Ketterle, Phys. Rev. Lett. {\bf 92}, 120403 (2004).

\bibitem{emery-kivelson}
V. J. Emery and S. A. Kivelson, Nature {\bf 374}, 434 (1995).

\bibitem{eth}
T. Stoferle, Henning Moritz, Kenneth Gunter, Michael Kohl, and Tilman Esslinger, Phys. Rev. Lett. {\bf 96}, 030401 (2006).

\bibitem{mit}
J. K. Chin, D. E. Miller, Y. Liu, C. Stan, W. Setiawan, C. Sanner, K. Xu, and W. Ketterle, Nature {\bf 443}, 961 (2006).

\bibitem{paiva}
T. Paiva, R. Scalettar, M. Randeria, and N. Trivedi, arXiv:0906.2141 (2009). 

\bibitem{Diener-Ho-lattice}
Roberto B. Diener and Tin-Lun Ho, Phys. Rev. Lett. {\bf 96}, 010402 (2006).

\bibitem{svistunov} E.~Burovski, N.~Prokof'ev, B.~Svistunov, and M.~Troyer, New J. Phys. {\bf 8}, 153 (2006).

\bibitem{definei} In 3D, $i$ at a given site can be thought of as a sum of the lattice indices in the 3 orthogonal directions. Note, the factor $(-1)^i$ then induces a relative $(-)$ sign between adjacent pairs of sites on a bipartite lattice e.g. a cubic lattice.

\bibitem{belkhir1}
L.~Belkhir and M.~Randeria, Phys. Rev. B {\bf 45}, 5087 (1992).

\bibitem{belkhir2}
L.~Belkhir and M.~Randeria, Phys. Rev. B {\bf 49}, 6829 (1994).

\bibitem{arun} A.~A.~Burkov and A.~Paramekanti, Phys. Rev. Lett. \textbf{100}, 255301 (2008). 

\bibitem{feedback} If one insists on treating $\mu$ and $\Delta$ on equal footing and wishes to feedback the fluctuations in the gap equation, the correct way to do so is to switch to an amplitude-phase representation for the fluctuations~\cite{rsr}. One then gets a different form for the fluctuation propagator corresponding to the gapless phase fluctuations and the validity of Goldstone's Theorem is ensured. However, this approach leads to an unphysical negative compressibility in the BEC limit in the continuum~\cite{rsr}.   

\bibitem{saddle point comment}
The saddle point condition for the integral (\ref{partition_function_Delta}) is the functional derivative $\delta S_\Delta[\Delta(x)] / \delta \Delta(x) = 0$, or $\delta S_\Delta[\Delta_\Q] / \delta \Delta_\Q =0$ for all $\Q$ in momentum space; this seems more general than the condition $dS_0 /d\Delta_0 = 0$ that we use and may suggest that we are using an approximate ``mean-field" saddle-point condition. 
This, however, is not the case: if we make an expansion of $S_\Delta$ around the static homogeneous saddle point $\Delta_0$ we get $S_\Delta(\Delta_0 + \eta_\Q) = S_0(\Delta_0) + (dS_0/d\Delta_0) \eta_0 + \sum_\Q \alpha_\Q \eta_\Q \eta_{-\Q}$, which shows that the only nontrivial saddle point equation for the {\it full} action is the one for $\Q = 0$ which corresponds to $dS_0/d\Delta_0 = 0$.


\bibitem{LeggettSF}
A. J. Leggett, Jn. of Stat. Phys. {\bf 93}, 927 (1998).

\bibitem{Taylor}
E. Taylor, A. Griffin, N. Fukushima, and Y. Ohashi,
Phys. Rev. A {\bf 74}, 063626 (2006).

\bibitem{garg_dmft} A. Garg, H. Krishnamurthy, and M. Randeria, Phys. Rev. B {\bf 72}, 24517 (2005).

\bibitem{trivedi95} N. Trivedi, and M. Randeria, Phys. Rev. Lett. {\bf 75}, 312 (1995).

\bibitem{randeria92} M. Randeria, N. Trivedi, A. Moreo, and R. Scalettar, Phys. Rev. Lett. {\bf 69}, 2001 (1992).

\bibitem{ohashi} H.~Tamaki, Y.~Ohashi and K.~Miyake, Phys. Rev. A {\bf 77}, 63616 (2008).

\bibitem{castellani}
A. Toschi, M. Capone, and C. Castellani, Phys. Rev. B {\bf 72}, 235118 (2005).

\bibitem{lutward}
J.~M.~Luttinger and J.~C.~Ward, Phys. Rev. {\bf 118}, 1417 (1960).

\end{thebibliography}
\end{document}